\documentclass[twocolumn]{aastex63}

\newcommand{\Msun}{$M_{\odot}\hspace{1mm}$}

\newcommand{\kms}{km s$^{-1}\hspace{1mm}$}

\newcommand{\lmass}{$\log\,(M_{*}/M_{\odot})$}

\def\lsim{~\rlap{$<$}{\lower 1.0ex\hbox{$\sim$}}}

\def\gsim{~\rlap{$>$}{\lower 1.0ex\hbox{$\sim$}}}

\shorttitle{NIR SN\,Ia }
\shortauthors{Johansson et al.}

\usepackage{ulem}

\begin{document}
\title{Near-IR Type Ia SN distances: host galaxy extinction and mass-step corrections revisited}

\correspondingauthor{J.~Johansson}
\email{joeljo@fysik.su.se}

\author[0000-0001-5975-290X]{J.~Johansson}
\affil{Oskar Klein Centre, Department of Physics, Stockholm University, AlbaNova, SE-10691 Stockholm, Sweden}

\author[0000-0002-2810-8764]{S.~B.~Cenko}
\affil{Astrophysics Science Division, NASA Goddard Space Flight Center, Mail Code 661, Greenbelt, MD 20771, USA}
\affil{Joint Space-Science Institute, University of Maryland, College Park, MD 20742, USA}

\author[0000-0003-2238-1572]{O.~D.~Fox}
\affil{Space Telescope Science Institute, 3700 San Martin Dr., Baltimore, MD 21218 USA}

\author[0000-0002-2376-6979]{S.~Dhawan}
\affil{Oskar Klein Centre, Department of Physics, Stockholm University, AlbaNova, SE-10691 Stockholm, Sweden}

\author[0000-0002-4163-4996]{A.~Goobar}
\affil{Oskar Klein Centre, Department of Physics, Stockholm University, AlbaNova, SE-10691 Stockholm, Sweden}

\author[0000-0002-7626-1181]{V.~Stanishev}
\affil{Department of Physics, Chemistry and Biology, IFM, Link{\"o}ping University, 581 83 Link{\"o}ping, Sweden}

\author{N.~Butler}
\affil{School of Earth and Space Exploration, Arizona State University, Tempe, AZ 85287, USA}

\author{W.~H.~Lee}
\affil{Instituto de Astronom{\'\i}a, Universidad Nacional Aut\'onoma de M\'exico, Apartado Postal 70-264, 04510 CDMX, M\'exico}

\author{A.~M.~Watson}
\affil{Instituto de Astronom{\'\i}a, Universidad Nacional Aut\'onoma de M\'exico, Apartado Postal 70-264, 04510 CDMX, M\'exico}

\author[0000-0002-4223-103X]{U.~C.~Fremling}
\affil{Division of Physics, Mathematics, and Astronomy, California Institute of Technology, Pasadena, CA 91125, USA}

\author[0000-0002-5619-4938]{M.~M.~Kasliwal}
\affil{Division of Physics, Mathematics and Astronomy, California Institute of Technology, Pasadena, CA 91125, USA}

\author[0000-0002-3389-0586]{P.~E.~Nugent}
\affil{Lawrence Berkeley National Laboratory, 1 Cyclotron Road, Berkeley, CA 94720, USA}
\affil{Department of Astronomy, University of California, Berkeley, Berkeley, CA 94720, USA}

\author[0000-0003-4743-1679]{T.~Petrushevska}
\affil{Centre for Astrophysics and Cosmology, University of Nova Gorica, Vipavska 11c, 5270 Ajdov\v{s}\u{c}ina, Slovenia}

\author[0000-0003-1546-6615]{J.~Sollerman}
\affil{The Oskar Klein Centre, Department of Astronomy, Stockholm University, AlbaNova, SE-10691 Stockholm, Sweden}

\author[0000-0003-1710-9339]{L.~Yan}
\affil{Caltech Optical Observatories, California Institute of Technology, Pasadena, CA 91125, USA}

\author[0000-0003-0035-6659]{J.~Burke}
\affil{Department of Physics, University of California, Santa Barbara, CA 93106-9530, USA}
\affil{Las Cumbres Observatory, 6740 Cortona Dr, Suite 102, Goleta, CA 93117-5575, USA}

\author[0000-0002-0832-2974]{G.~Hosseinzadeh}
\affil{Center for Astrophysics \textbar{} Harvard \& Smithsonian, 60 Garden Street, Cambridge, MA 02138-1516, USA}

\author[0000-0003-4253-656X]{D.~A.~Howell}
\affil{Department of Physics, University of Cal ifornia, Santa Barbara, CA 93106-9530, USA}
\affil{Las Cumbres Observatory, 6740 Cortona Dr, Suite 102, Goleta, CA 93117-5575, USA}

\author[0000-0001-5807-7893]{C.~McCully}
\affil{Department of Physics, University of California, Santa Barbara, CA 93106-9530, USA}
\affil{Las Cumbres Observatory, 6740 Cortona Dr, Suite 102, Goleta, CA 93117-5575, USA}

\author[0000-0001-8818-0795]{S.~Valenti}
\affil{Department of Physics and Astronomy, University of California, 1 Shields Avenue, Davis, CA 95616-5270, USA}

\begin{abstract}
We present optical and near-infrared (NIR, $YJH$-band) observations of 42 Type Ia supernovae (SNe Ia) discovered by the untargeted intermediate Palomar Transient Factory (iPTF) survey. This new data-set 
covers a broad range of redshifts and host galaxy stellar masses, compared to previous SN\,Ia efforts in the NIR.
We construct a sample, using also literature data at optical and NIR wavelengths, to examine claimed correlations between the host stellar masses and the Hubble diagram residuals. The SN magnitudes are corrected for host galaxy extinction using either a global total-to-selective extinction ratio, $R_V$=2.0 for all SNe, or a best-fit $R_V$ for each SN individually.
Unlike previous studies which were based on a narrower range in host stellar mass, we do not find evidence for a "mass-step", between the color- and stretch-corrected peak $J$ and $H$ magnitudes for galaxies below and above $\log(M_{*}/M_{\odot}) = 10$. However, the mass-step remains significant ($3\sigma$) at optical wavelengths ($g,r,i$) when using a global $R_V$, but vanishes
when each SN is corrected using their individual best-fit $R_V$. 
Our study confirms the benefits of the NIR SN\,Ia distance estimates, as these are largely exempted from the empirical corrections dominating the systematic uncertainties in the optical.   
\end{abstract}


\section{Introduction} \label{sec:intro}

Since the initial standardization of Type Ia supernova (SN~Ia) peak luminosities was employed in the discovery of the accelerated expansion of the Universe \citep{riess1998,perlmutter1999}, estimates of the local value of the Hubble constant from SNe \citep[$H_0$][]{riess2019} are in tension with the value inferred from the early universe \citep{planck2020}. This tension is a possible sign of new physics or unresolved sources of systematic uncertainty. 

Significant work has gone into understanding how to more precisely standardise SNe~Ia as distance indicators at optical (visible) wavelengths. The SN~Ia optical peak brightness is corrected for lightcurve shape \citep{phillips1993} and color \citep{tripp1998}, and there are now several more elaborated prescriptions for optimising these standardisation procedures \citep[see, e.g.,][]{guy2007,burns2011,mandel2011}. More recently, additional correction terms aiming at further improving the SN~Ia standard candle have also been proposed.
One such term accounts for the dependence of the SN~Ia luminosity on its host galaxy properties, e.g. stellar mass \citep{hamuy1995, sullivan2003,2010ApJ...722..566L,childress2013,  betoule2014, 2017ApJ...850..135U, 2018ApJ...859..101S, 2020MNRAS.495.4040W,kelsey2021}. These studies all uncover, to various degrees of significance, a ``mass step" in the data: after light-curve standardisation,
SNe in high-mass galaxies are more luminous than those exploding in low-mass galaxies.

The origin of this mass step is poorly understood, with possible explanations suggesting that it is due to dust in the host galaxies \citep{broutscolnic2021}.

Near-infrared (NIR; 1 $< \lambda < 2.5~\mu$m) observations offer many advantages for standardising SNe Ia \citep{elias1985,meikle2000}. Not only is the NIR less prone to extinction from dust, but SNe~Ia are more naturally standard candles at these wavelengths, requiring no or significantly smaller corrections to their peak luminosity to yield similar precision as compared to the optical
\citep{2004ApJ...602L..81K,2008ApJ...689..377W,2009ApJ...704..629M,burns2011,dhawan2018a,burns2018,avelino2019}.
Theoretical models further corroborate these observations \citep{kasen2006,2015MNRAS.448.2766B}. 

There are already upcoming datasets \citep[e.g. CSP-II, Sweetspot and RAISIN;][]{Phillips2019,ponder2020,2013IAUS..281....1K}, and ongoing \citep[e.g. SIRAH, DEHVILS and VEILS;][]{sirah} and future SN~Ia programs \citep[e.g. with the Nancy Grace Roman Space Telescope;][]{2018ApJ...867...23H} aiming to take advantage of these properties of SNe~Ia and use NIR observations to study dark energy.
In this contexts, NIR observations of SNe~Ia in the nearby Hubble flow ($z \gsim 0.03$)
are extremely valuable cosmological tools both as a Hubble flow rung of the local distance ladder and as a low-$z$ "anchor" sample to measure dark energy properties. 

However, as \citet{burns2018} point out, there is a deficit of SNe in low-mass hosts in the current SN~Ia NIR data-set and observing an unbiased sample of SNe~Ia in the nearby Hubble flow is crucial to test the impact of SN~Ia systematics, e.g. extinction from host galaxy dust, on the inferred value of $H_0$ \citep{dhawan2018a,burns2018}.
Moreover, recent works have also claimed evidence for a mass step in the NIR as well \citep{uddin2020,ponder2020}. If indeed present and not accounted for, it will introduce further systematic uncertainties in the NIR SN~Ia cosmological analyses.

The main goal of this work is to obtain optical and NIR light curves of an unbiased sample of SNe~Ia in the nearby Hubble flow, and together with data from the literature to examine the impact of the host galaxy extinction determination on the claimed correlations between the host stellar masses and the NIR Hubble diagram residuals.

Here we present optical and NIR observations of a new sample of 42 SNe~Ia with redshifts out to $z \sim 0.12$ and containing 12 SNe in hosts with masses below \lmass =10.

Section \ref{sec:sample} presents our sample.  Section \ref{sec:observations} describes our observations.  Section \ref{sec:analysis} presents our analysis techniques, including spectroscopic classification, light-curve fitting, derivation of the NIR Hubble diagram, and correlations with the host galaxy stellar mass.  Section \ref{sec:discussion} discusses of the results, and Section \ref{sec:conclusion} provides our conclusion.

Throughout this paper we assume flat $\Lambda$CDM cosmological model with $\Omega_M$ = 0.27 and Hubble constant $H_0$ = 73.2~km\,s$^{-1}$\,Mpc$^{-1}$ from \citet{burns2018}.

\vspace{12pt}
\section{Supernova sample}
\label{sec:sample}
This work presents 42 new SNe Ia discovered with the intermediate Palomar Transient Factory \citep[iPTF;][]{rau2009}. We chose targets spanning a wide range of redshifts and host galaxy environments, and acquired optical and NIR follow-up observations for targets with early iPTF detection and classification.  These observations are described in more detail in Section \ref{sec:observations}.  

\begin{figure}
    \includegraphics[angle=0,width=1.0\columnwidth]{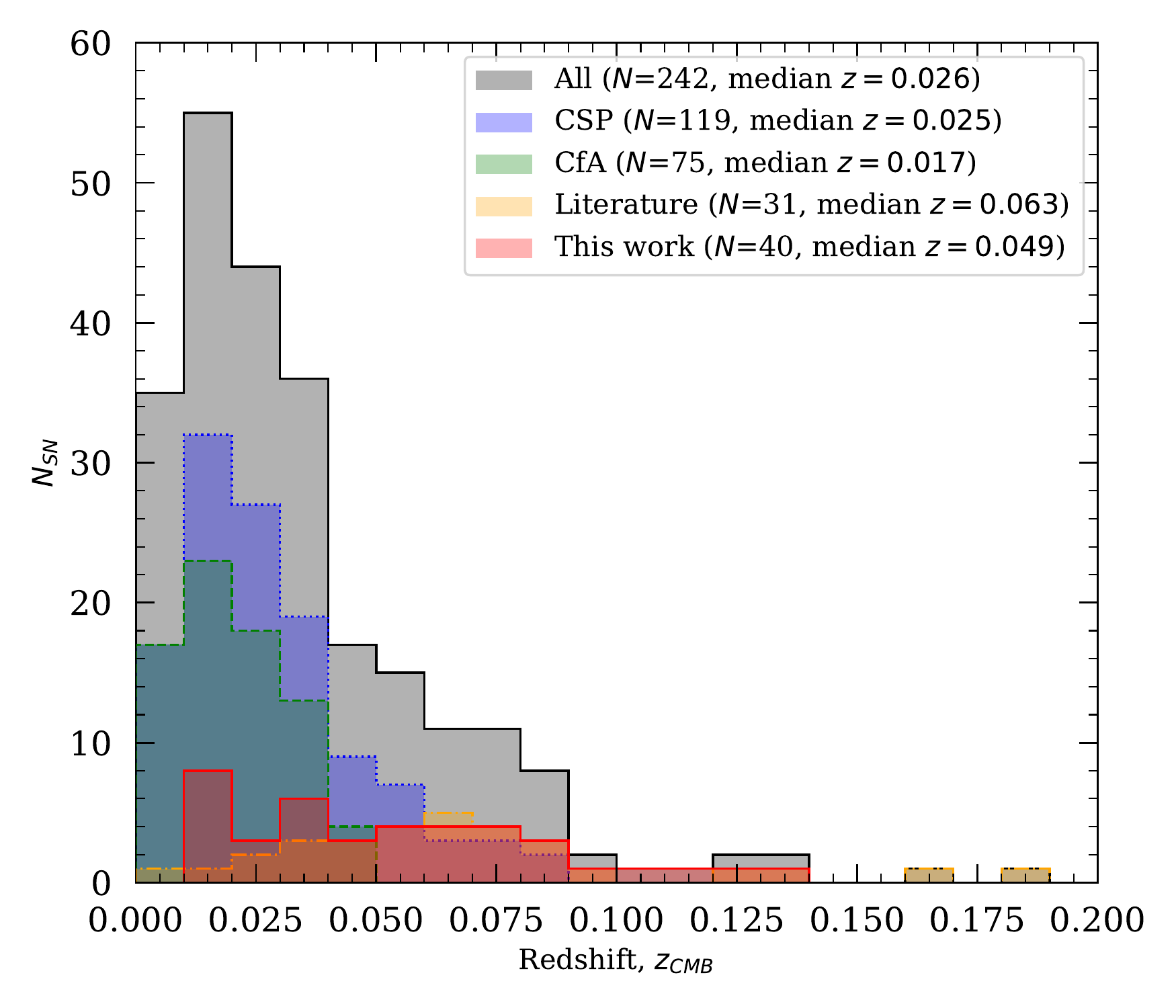}
    \caption{Redshift distribution of the SNe used in the analysis. Note that peculiar SNe~Ia are not included here, as well as SNe lacking optical lightcurves. The total number of unique SNe with both optical and NIR lightcurves amounts to 242. 
    }
    \label{fig:redshift_samples}
\end{figure}

\begin{figure}
    \includegraphics[angle=0,width=1.0\columnwidth]{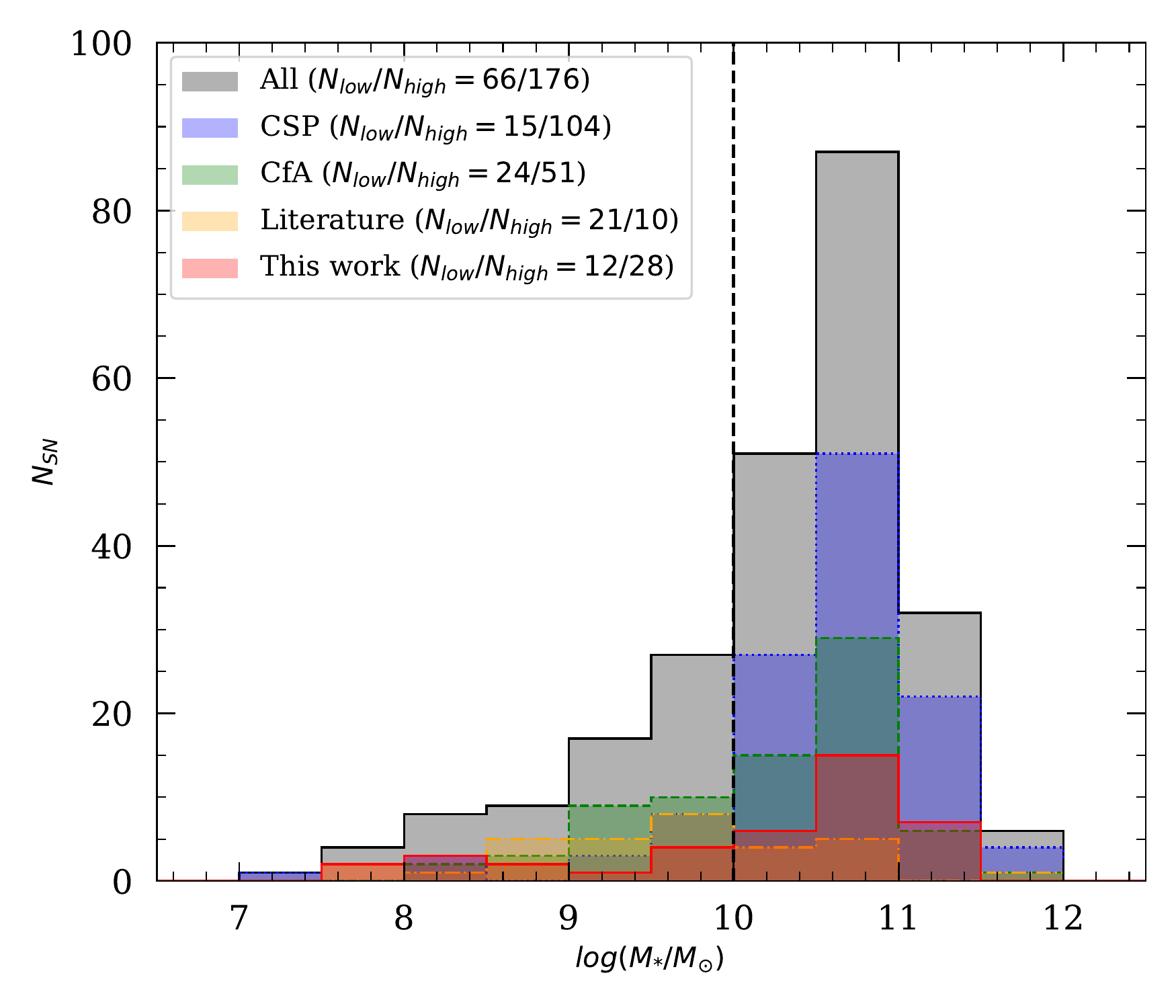}
    \caption{Host galaxy mass distribution of the SNe used in the analysis.}
    \label{fig:hostmasses_samples}
\end{figure}

For our analysis, we also include SNe Ia from the literature having both optical and NIR light curves, which we describe briefly here and summarize in Figures~\ref{fig:redshift_samples} and \ref{fig:hostmasses_samples}.  The final photometry of the first stage of the Carnegie Supernova Project (CSP-I) are presented in \citet{2017AJ....154..211K}. Their sample consists of 120 SNe with NIR coverage,  $z$=0.0037 to 0.0835.  
CfAIR2 \citep{friedman2015} is a sample of NIR light curves for 94 SNe~Ia obtained with the 1.3m Peters Automated InfraRed Imaging TELescope (PAIRITEL) between 2005-2011. \citet{baronenugent2012} present $J$ and $H$-band lightcurves of 12 SNe~Ia discovered by PTF in the redshift range $0.03 < z < 0.08$. This data was re-analysed by \citet{stanishev2018}, including optical lightcurves. \citet{stanishev2018} add 16 more SNe with NIR data in the redshift range $z$=0.037 to 0.183. Furthermore, we include the 6 SNe with UV, optical and NIR lightcurves in \citet{Amanullah2015}. Note that some of the supernovae were observed by, e.g., both CSP and CfA \citep[see][for a comparison]{friedman2015}, and the total sample size in Figures \ref{fig:redshift_samples} and \ref{fig:hostmasses_samples} refers to the number of unique SNe.

\section{Observations}\label{sec:observations}
The follow-up observations were obtained with several different facilities, which are described in the following sections. For each instrument used, deep reference images were obtained after the supernova emission had faded away. The reference images were subtracted from the science images in order to facilitate the photometry of the SNe, which can otherwise be affected by the light of the host galaxy. Image subtraction was in most cases performed as part of the reduction pipelines, which all utilize implementation of the convolution algorithms presented in \citet{alard1998}.

\subsection{Optical data}
During the intermediate Palomar Transient Factory (iPTF) survey, the Palomar 48-inch ($P48$) telescope typically delivered $g$ and $R$-band images.
The $P48$ image reduction is described by \citet{laher2014}, while the PTF photometric calibration and the photometric system are discussed by \citet{ofek2012}.

Optical follow-up observations were collected using the Palomar 60-inch telescope (P60, $BVgriz$ filters), the 2.56 m Nordic Optical Telescope (NOT) and the Las Cumbres Observatory (LCO) in $UBVRI$ and/or $griz$-bands.
The P60 data were reduced using an automated pipeline \citep{cenko2006}, calibrated against SDSS and the reference images subtracted using \texttt{FPipe} \citep{fremling2016}.
Similarly, the NOT data were reduced with standard IRAF routines using the \texttt{QUBA} pipeline \citep{valenti2011}, calibrated to the Landolt system through observations of standard stars and SDSS stars in the field.
LCOGT data were reduced using \texttt{lcogtsnpipe} \citep{valenti2016} by performing PSF-fitting photometry. Zeropoints for images in the $UBVRI$ filters were calculated from Landolt standard fields \citep{landolt1992} taken on the same night by the same telescope. For images in the $griz$ filter set, zeropints were calculated using SDSS magnitudes of stars in the same field as the object.

\subsection{Near-IR observations}
For 37 out of 42 SNe in our sample, we acquired follow-up observations using the Reionization and Transients InfraRed camera (RATIR).  RATIR is a six band simultaneous optical and NIR imager ($riZYJH$-bands) mounted on the autonomous  1.5~m Harold  L.   Johnson Telescope  at the  Observatorio Astron\'{o}mico  Nacional  on  Sierra  San Pedro  M\'{a}rtir  in  Baja California, Mexico
\citep{2012SPIE.8446E..10B,2012SPIE.8444E..5LW,2012SPIE.8453E..2SK,2012SPIE.8453E..1OF}.

Typical observations include a series of 80-s exposures in the $ri$-bands and 60-s exposures in the $ZYJH$~bands, with dithering between exposures. The fixed IR filters of RATIR cover half of their respective detectors, automatically providing off-target IR sky exposures while the target is observed in the neighbouring filter. Master IR sky frames are created from a median stack of off-target images in each IR filter. No off-target sky frames were obtained on the optical CCDs, but the small galaxy sizes and sufficient dithering allowed for a sky frame to be created from a median stack of selected images in each filter that did not contain either a bright star or extended host galaxy. 

\begin{figure*}
    \centering
    \includegraphics[width=\textwidth]{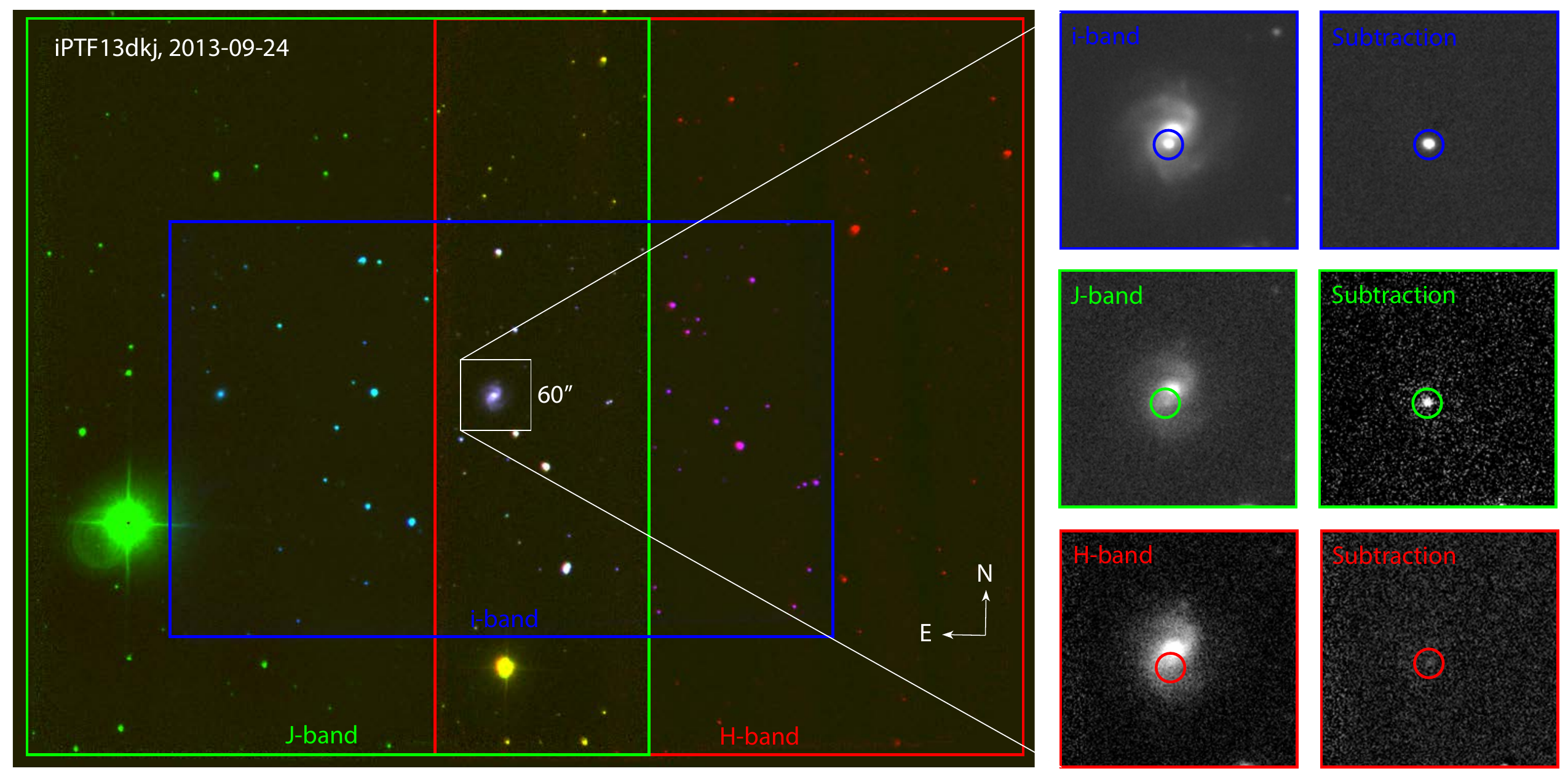}
    \caption{Example of typical RATIR observations for iPTF13dkj (at $z=0.036$) from 2013-09-24. For the RGB composite (left panel), the blue, green and red insets show $i$, $J$ and $H$-band images, respectively. The middle and right panels show the 60\arcsec $\times$ 60\arcsec region centered on the SN and host galaxy.
    }
    \label{fig:ratir}
\end{figure*}

Flat-field frames consist of evening sky exposures. Given the lack of a cold shutter in RATIR’s design, IR dark frames are not available. Laboratory testing, however, confirms that the dark current is negligible in both IR detectors \citep{fox12}.  Bias  subtraction and  twilight flat division  are performed
using algorithms written in {\sc python}, image alignment is conducted by astrometry.net \citep{2010AJ....139.1782L} and image co-addition is
achieved using {\sc swarp} \citep{2010ascl.soft10068B}.  Figure \ref{fig:ratir} shows a typical set of images, where blue, green and red shows the field-of-view for $i$, $J$ and $H$-band frames, respectively.

For seven SNe~Ia in our sample, $J$ and $H$-band observations were also obtained using other facilities, such as HAWK-I on the 8m Very Large Telescope (VLT) (for iPTF14bbr, 14ddi, 14deb, 14eje and 14fww), VIRCAM on the 4m VISTA telescope (iPTF14fpb) and WIRC on the Palomar 200-inch telescope (iPTF14gnl).
These observations were processed with the corresponding instrument reduction pipelines.

\subsection{Image Subtraction}
Image subtraction was performed utilizing the High Order Transform of PSF ANd Template Subtraction \citep[\texttt{HOTPANTS};][]{Becker2015}. Point sources were selected across the field-of-view (FOV) to calculate the point-spread function (PSF) in each image, either based on classification from SDSS or through manual inspection. Given the relative paucity of bright point sources in most fields (particularly in the NIR), the PSF was held fixed across the FOV. 

The calculated PSFs were utilized to perform PSF-matched photometry on the resulting subtracted images, yielding measurements of the instrumental magnitude of the supernova in each epoch:
\begin{equation}
	m_{f,\mathrm{inst}} = -2.5 \log_{10} \frac{\rm ADU}{t_{\mathrm{exp}}}
\end{equation}
Uncertainties and upper limits were determined by inserting false sources of varying brightness into the RATIR images and repeating the identical process of image subtraction and PSF-matched photometry. 

\subsection{Photometric Calibration}
Photometric calibration of the RATIR data was performed following the process outlined in \citet{ofek2012}. To calculate color and illumination terms, we selected fields with coverage from both SDSS (optical) and UKIDSS (NIR), and obtained photometry for stars (i.e., objects classified as point sources in SDSS) with $r$-band magnitudes between 14 and 18 (with additional flagging for saturation). We measured instrumental magnitudes via PSF-matched photometry for these calibration stars as above.

As a first pass, we calculate a zero-point for each image with no additional corrections (e.g., color and illumination terms). We removed nights with large scatter in the zeropoint (RMS $\geq 0.10$\,mag) or individual stars that were clear outliers in the fits (determined via visual inspection). 

We then performed a least squares fit using the remaining nights/stars to the following equation for each filter $f$: 
\begin{eqnarray}
	m_{f} & = & m_{f,\mathrm{inst}} + ZP + CT_\mathrm{f,i,j} \cdot (m_\mathrm{i} - m_\mathrm{j}) + C_\mathrm{illum} 
\label{eqn:calibration}
\end{eqnarray}
where $ZP$ is the zero-point, $CT_{\rm f,i,j}$ is the color term, $m_{\rm i}$ and $m_{\rm j}$ are the filters used for the color correction, and $C_{\rm illum}$ is an illumination correction term accounting for PSF variations depending on the position on the detector.

The color and illumination terms were held fixed for all observations in a given filter, while the zero-point term was allowed to vary freely in each image. The resulting best-fit color terms and zero-point RMS are shown in Table~\ref{tab:calibration}. The zero-point RMS is typically $\sim$0.03 mag for the RATIR $r$ to $H$-band.

For fields with SDSS and UKIDSS coverage, calibrated supernova magnitudes were calculated using Equation~\ref{eqn:calibration}. For fields lacking SDSS coverage, we used photometry from Pan-STARRS1 Data Release 2 \citep{magnier2020}, which is in a photometric system close to SDSS \citep{Tonry_2012}. For fields lacking UKIDSS coverage we used 2MASS \citep{skrutskie2006} and the transformation from \citet{Hodgkin2009} to calibrate the $Y$-band RATIR data ($Y = J + 0.50 \times (J - H) + 0.08$). 

\begin{table}
    \centering
  \caption{RATIR photometric calibration}
  \label{tab:calibration}
  \begin{tabular}{@{}llllll@{}}
    \hline \hline
Filter & Color term & Color& ZP RMS & Limiting  \\
	& $C_{f,ij}$  & ($i-j$) & [mag] & [mag] \\
\hline
$r$ & 0.009 & ($r - i$) & 0.031 & 21.54 \\
$i$ & 0.030 & ($r - i$) & 0.025 & 21.52 \\
$z$ & -0.048 & ($i - z$) & 0.032 & 20.80 \\
$Y$ & 0.046 & ($Y - J$) & 0.031 & 19.81 \\
$J$ & 0.057 & ($J - H$) & 0.026 & 18.94 \\
$H$ & -0.054 & ($J - H$) & 0.032 &  18.27 \\
    \hline \hline
    \end{tabular}
\end{table}

\section{Analysis}\label{sec:analysis}
Some of the SNe presented here have previously been published in separate papers:
\begin{itemize}
\item Optical and NIR light curves and spectra of iPTF13abc (SN~2013bh) were presented and analysed in \citet{silverman2013}. It is a near identical twin to the peculiar Ia SN~2000cx.
\item UV, optical and NIR light curves and spectra of iPTF13asv (SN\,2013cv) were presented in \citet{cao2016} and has additional $H$-band photometry in \citet{weyant2018}. iPTF13asv shows low expansion velocities and persistent carbon absorption features after the maximum, both of which are commonly seen in super-Chandrasekhar events, although its light curve shape and sharp secondary near-IR peak resemble characteristic features of normal SNe~Ia.
\item Optical light curves and high-resolution spectra of iPTF13dge were presented in \citet{ferretti2016}, and NIR light curves in \citet{weyant2018}. The light curves are compatible with a normal SN Ia with little reddening, and no definite time-variability could be detected in any absorption feature of iPTF13dge.
\item UV, optical and NIR observations of iPTF13ebh from the CSP-II collaboration were presented in \citet{hsiao2015}. iPTF13ebh can be categorized as a "transitional" event, on the fast-declining end of normal SNe Ia, showing NIR spectroscopic properties that are distinct from both the normal and subluminous/91bg-like classes. 
\item iPTF14atg is a subluminous peculiar SN similar to SN\,2002es. It displayed strong, declining ultraviolet emission shortly after explosion. Spectra together with UV, optical and NIR photometry have been extensively analysed in \citet{cao2015,kromer2016}. 
\item UV and optical photometry and spectra of the 1999aa-like SN iPTF14bdn were presented in \citet{smitka2015}. 
\item iPTF16abc was analyzed by \citet{miller2018}, \citet{Ferretti2017} and \citet{dhawan2018b}. The rapid, near-linear rise, the non-evolving blue colors, and strong absorption from ionized carbon, are interpreted to be the result of either vigorous mixing of radioactive-Ni in the SN ejecta, or ejecta interaction with diffuse material, or a combination of the two. 
\item iPTF17lf was reddened, spectroscopically normal SN\,Ia, discovered during a wide-area (2000 deg$^2$) $g$ and $I$-band survey for "cool transients" as part of a two month extension of iPTF \citep{adams2018}. 
\end{itemize}
For the other SNe included in our sample (except iPTF14ale, which has no spectroscopic classification), we run the SuperNova IDentification code \citep[SNID][]{blondin2007} on the spectra (to be presented in a separate paper). For SNe iPTF13s, iPTF13ddg, iPTF13efe, iPTF14bpz, iPTF14fpb we rely on redshift estimates based on the SN spectral features using \texttt{SNID}. Furthermore, for iPTF13anh, iPTF13asv, iPTF13azs, iPTF13crp and iPTF13dkx, we determine the redshifts from narrow host galaxy lines in the SN spectra.

\begin{figure}[htp]
    \centering
    \includegraphics[width=0.5\columnwidth]{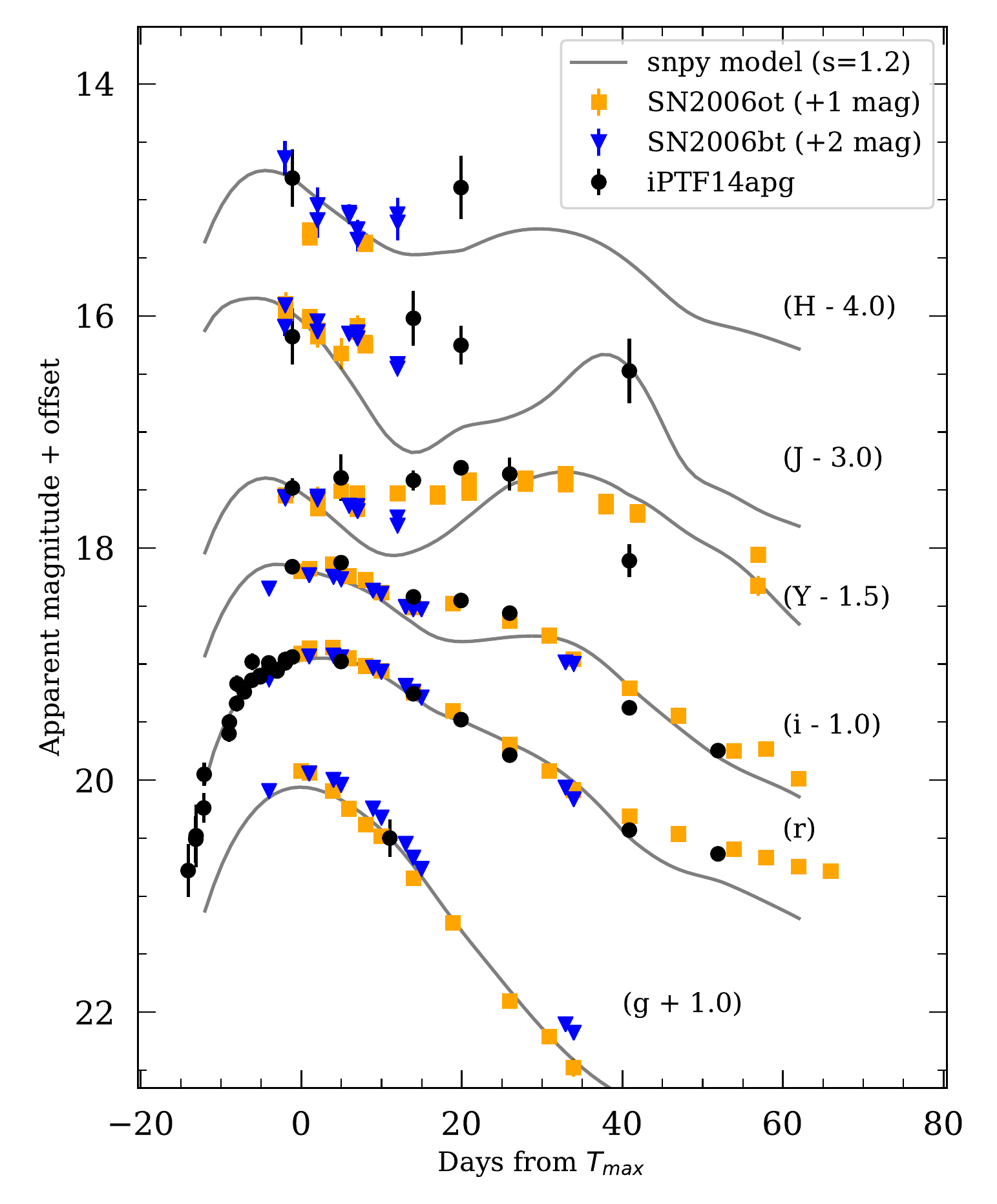}
    \caption{Black symbols show our observed light curves of iPTF14apg. The \texttt{SNooPy max\_model} fails to accurately fit the light curves, shown in gray lines (based on the $r$-band stretch, and offset to match the peak magnitudes). In blue and orange are the lightcurves of peculiar Ia SNe 2006bt and 2006, respectively. 
    }
    \label{fig:14apg_lc}
\end{figure}

For our single spectrum of iPTF14apg, observed 5 days before peak brightness, \texttt{SNID} gives a best match to SN\,2004dt at $z=0.088\pm0.004$, consistent with the spectroscopic redshift of the nearest galaxy. Among the top matches are also SNe 2006ot and 2006bt \citep{foley2010}, which are peculiar Ia SNe excluded from the Hubble diagram analysis \citet{burns2018,uddin2020}. A direct comparison of the light curves of iPTF14apg to those of SNe 2006ot and 2006bt (see Fig.~\ref{fig:14apg_lc}) strengthens this classification.

\subsection{Host galaxies}
\label{sec:hostmasses}
\begin{figure*}[!t]
    \centering
    \includegraphics[width=0.90\textwidth]{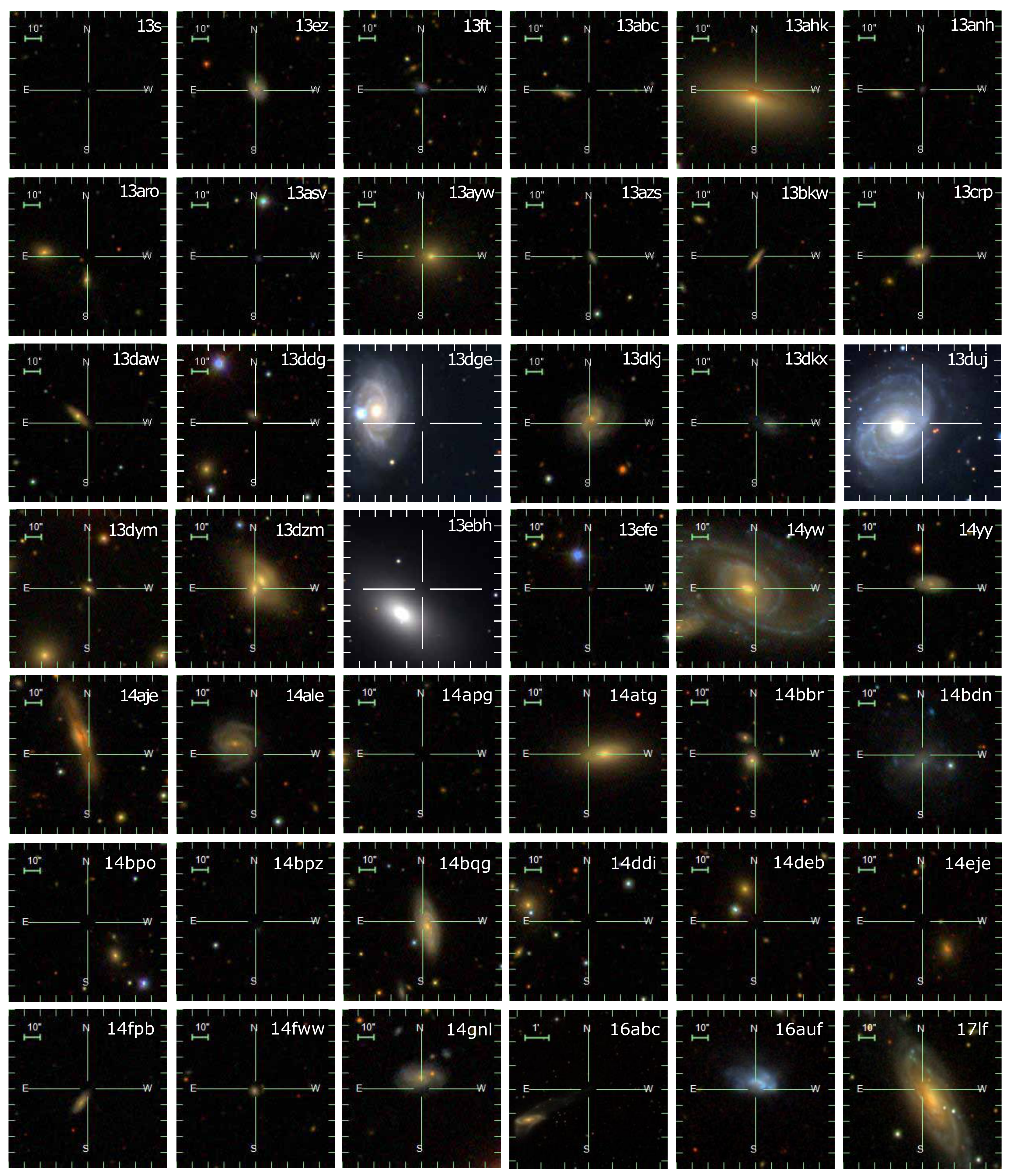}
    \caption{Cutout stamps from SDSS (and PS1) showing the SNe and host galaxies in our sample.}
    \label{fig:patches}
\end{figure*}

Figure~\ref{fig:patches} shows cut-out images from the SDSS and PanStarrs surveys, centered on the SN positions. Most SNe can easily be associated with their hosts, while some cases are ambiguous, including: 
\begin{itemize}
\item iPTF14apg: nearest galaxy is SDSS J123758.69+082301.5 with a
spectroscopic redshift $z$=0.08717, separated by 51\arcsec, corresponding to a projected distance of 79.4 kpc.
\item iPTF14bpo: nearest galaxy is SDSS J171429.74+310905.0 with a spectroscopic redshift $z$=0.07847, separated by 27\arcsec, corresponding to a projected distance 38.9 kpc.
\item iPTF14ddi: nearest galaxy is SDSS J171036.45+313945.0 with a spectroscopic redshift $z$=0.08133, separated by 40\arcsec, corresponding to a projected distance 59.2 kpc.
\end{itemize}
For the literature sample, we note that SNe PTF10hmv, PTF10nlg and PTF10qyx from \citet{baronenugent2012} have ambiguous hosts. 

We estimate the host galaxy stellar mass, $M_{*}$, using the relationship published in \citet{taylor2011},
\begin{equation}
    \log(M_{*}/M_{\odot}) = 1.15 + 0.7(m_g - m_i) - 0.4 M_i .
\end{equation}
We use $g$ and $i$-band magnitudes from SDSS (or PanStarrs when no SDSS photometry was available), corrected for the Milky Way (MW) extinction. $M_i$ is the absolute magnitude in the $i$-band. Table~\ref{tab:sne} lists the redshifts and coordinates of the SNe in our sample, together with their likely host galaxies and our estimates of the host galaxy stellar mass.

\begin{figure}[!t]
    \centering
    \includegraphics[angle=0,width=1.0\columnwidth]{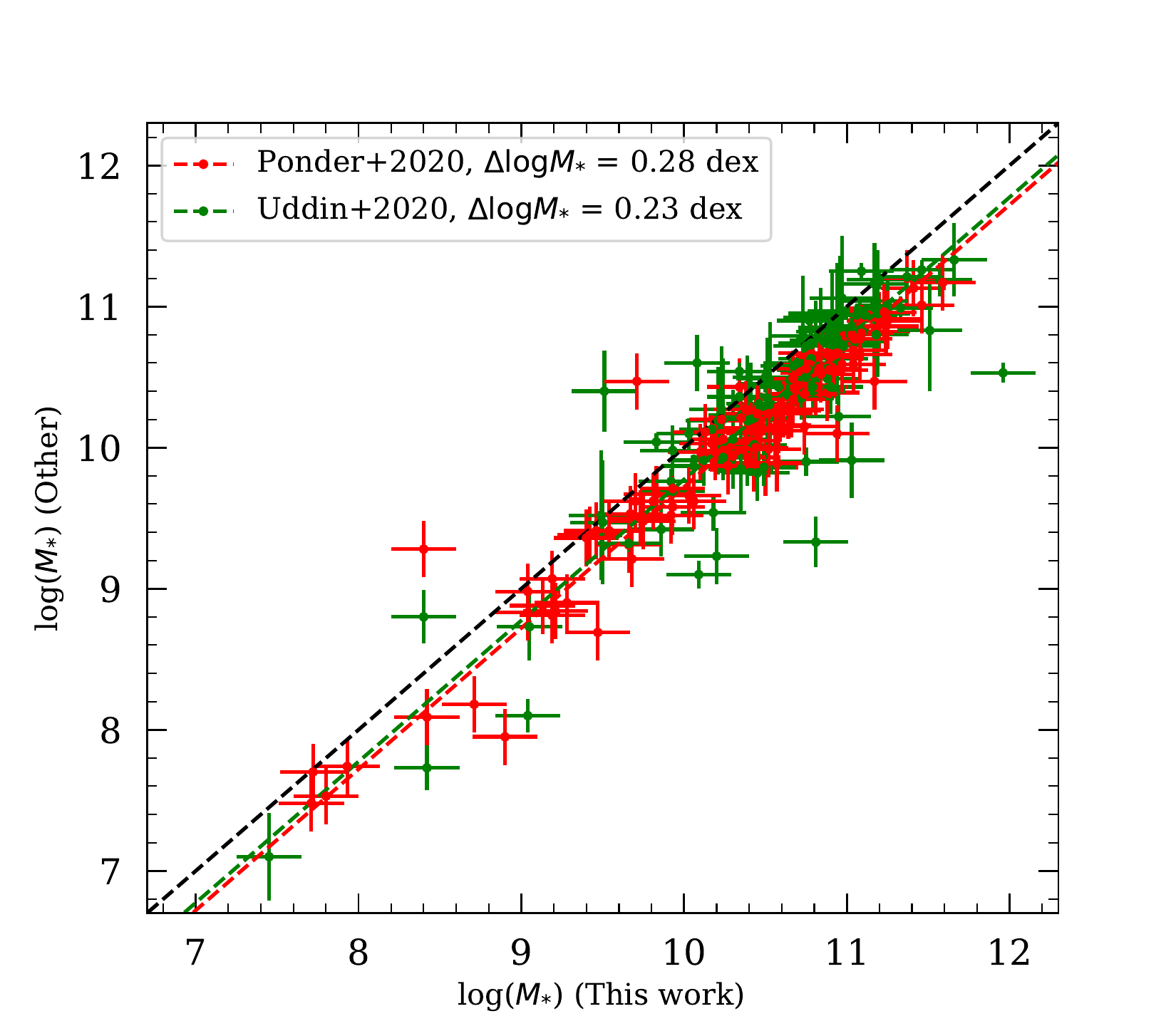}
    \caption{Host galaxy mass estimates from this work, compared to the SNe in common with \citet{ponder2020} and \cite{uddin2020}.}
    \label{fig:hostmasses_comparison}
\end{figure}

Our mass estimates are consistent with those of \citep{neill09,chang2015,burns2018}, but systematically higher by $\sim$0.2-0.3 dex than the estimates from \citet{ponder2020} and \citet{uddin2020}, who employ a more sophisticated SED fitting (Fig.~\ref{fig:hostmasses_comparison}). However, for consistency when comparing stellar masses between our sample and the CSP, CfA and literature sample, we choose to use our estimates for the combined analysis.

\subsection{Light curve and host galaxy extinction fitting}
\begin{figure}[!t]
    \centering
    \includegraphics[width=\columnwidth]{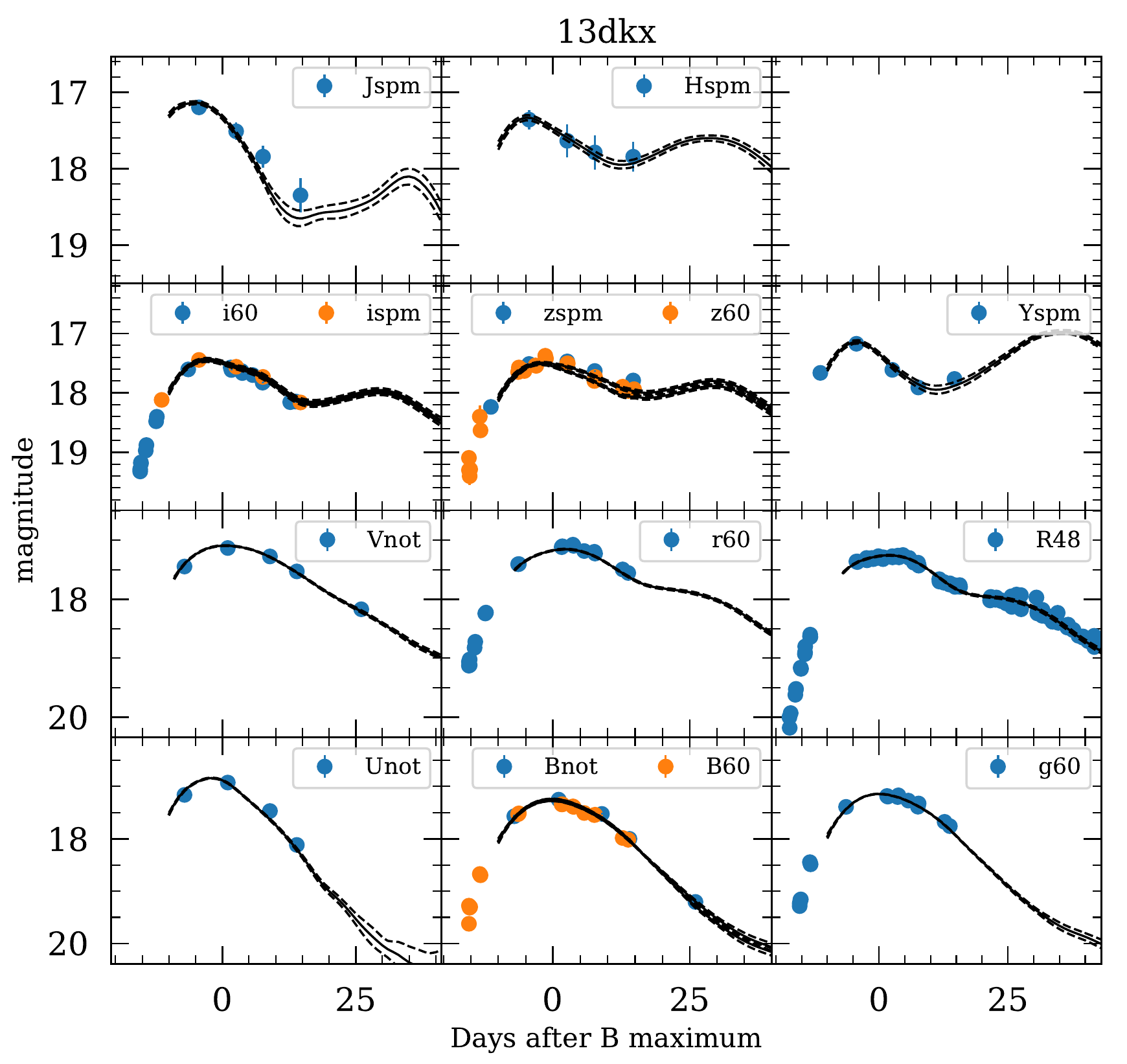}
    \includegraphics[width=\columnwidth]{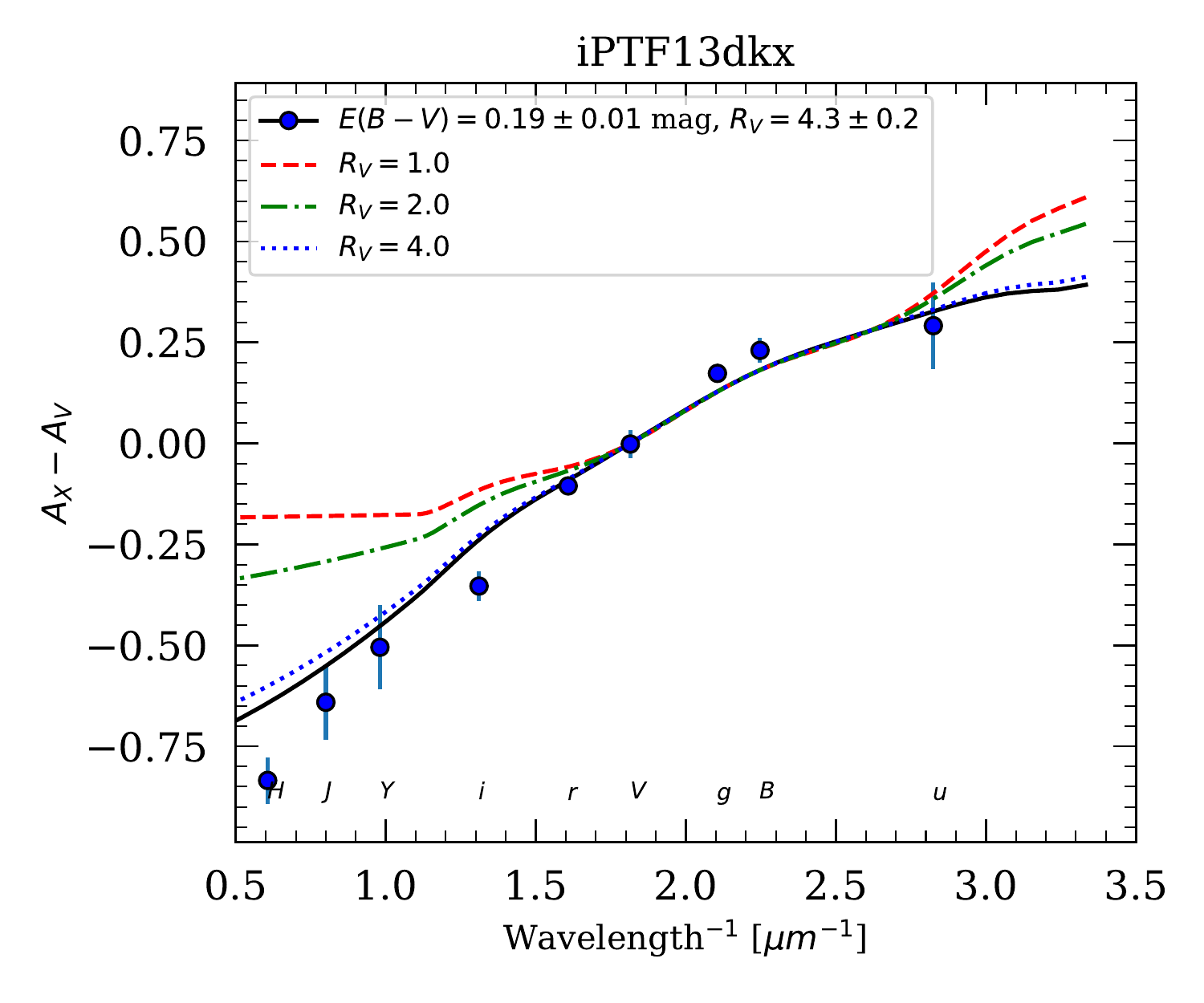}
    \caption{Upper panel shows the \texttt{max\_model} 
    fit to iPTF13dkx, a representative SN from our sample at "moderate" redshift and NIR coverage around peak brightness. Lower panel shows the inferred color excess (normalized with respect to $V-$band) and the best-fit extinction parameters. 
    }
    \label{fig:13dkx_lc}
\end{figure}
We use the \texttt{SNooPy} light curve fitting package developed for the CSP sample \citep{burns2011,burns2014,burns2018} to analyze the light curves the SNe in our sample, including the light curve of the literature sample.
To find the time of maximum, $T_{\rm max}$, and color-stretch parameter, $s_{BV}$, and the observed rest-frame peak magnitudes\footnote{The MW extinction is included in the fitted model and the derived magnitudes are corrected for it.} of the SNe, the \texttt{SNooPy max\_model} was fitted to the light curves. An example fit is shown in  the upper panel of Figure~\ref{fig:13dkx_lc} and the derived light curve parameters are given in Tables \ref{tab:peakmags} and \ref{tab:lcfits}.

To derive the host galaxy extinction we use the more elaborated \texttt{color\_model}. This model allows to fit for the host galaxy extinction taking into account the dependence of SN~Ia intrinsic colors on $s_{BV}$ \citep{burns2014}. It uses parametrized dust extinction laws to calculate the total-to-selective extinction ratio $R_X$ in any filter $X$, as a function of $R_V$ and $E(B-V)_\mathrm{host}$ by the means of synthetic photometry \citep[see][]{burns2011}. 
As $R_V$ controls the wavelength dependence of the extinction and the host-galaxy color excess $E(B-V)_\mathrm{host}$ the amount of the extinction, with observations over a broad range of frequencies it is in principle possible to fit independently for $R_V$ and $E(B-V)_\mathrm{host}$, which are otherwise correlated. In our analysis we used \citet{ccm1989} and \citet{1994ApJ...422..158O} extinction law. For full details on the \texttt{color\_model} the reader is referred to \citet{burns2014,burns2018}.

We performed two fits for the extinction. First, $R_V=2.0$ was assumed for all SN hosts and only $E(B-V)_\mathrm{host}$  was fitted. The value $R_V=2$ corresponds to our sample average (weighted average $R_V = 1.9$, $\sigma_{R_V}=0.8$) and is close to values commonly found in many SN~Ia cosmological analyses, which commonly employ single $R_V$. Second, both $E(B-V)_\textrm{host}$ and $R_V$ were fitted. This is possible because SNe Ia show a small intrinsic color dispersion across optical to NIR bands and the wavelength leverage provided by including NIR observations. Nevertheless, when $E(B-V)_\mathrm{host}$ approaches zero (or rather the level of scatter in the intrinsic color, $\sigma_{E(B-V)} \sim 0.06$ mag), the leverage to get meaningful constraints on $R_V$ decreases. The results from the second fit are shown in Table~\ref{tab:lcfits}. Figure~\ref{fig:13dkx_lc} lower panel shows an example of the inferred color excess and the best-fit extinction parameters. 

It is a long-standing issue that SN analyses have yielded "unusually"\footnote{it may be that the MW average of $R_V=3.1$ is unusually high} low $R_V$ values. This is seen both when minimizing the Hubble residuals using a global $R_V$ for cosmological samples and for detailed studies of individual, highly-extinguished SNe \citep[e.g. SNe 2006X and 2014J;][]{burns2014,Amanullah2014}.
We stress that we only use the observed colors to constrain $E(B-V)_\mathrm{host}$ and $R_V$,  since determining extinction by minimizing Hubble residuals can lead to a bias \citep{burns2018,uddin2020}.

\subsection{NIR Hubble diagram}\label{sec:hubblediagram}
To construct the Hubble diagrams, the distance modulus for filter $X$, $\mu_X$, was computed as:
\begin{equation}
\label{eq:reddening_model}
    \mu_{X}  =  m_{X} - P_{X}^{N}(s_{BV}-1) - R_{X,BV}E(B-V)_\mathrm{host}, 
\end{equation}
where $P_{X}^{N}(s_{BV}-1)$ is the 2-nd order polynomial luminosity-decline-rate relation from \citet{burns2018} and $R_{X,BV}$ is
the total-to-selective absorption coefficient for filter $X$ computed from $R_V$ and $E(B-V)_\mathrm{host}$. Here, we impose that $R_V > 0$ and do not correct for dust extinction  objects with $E(B-V)_\mathrm{host}<0$, i.e. intrinsically blue objects. 

\begin{figure}[!t]
    \centering
    \includegraphics[angle=0,width=1.0\columnwidth]{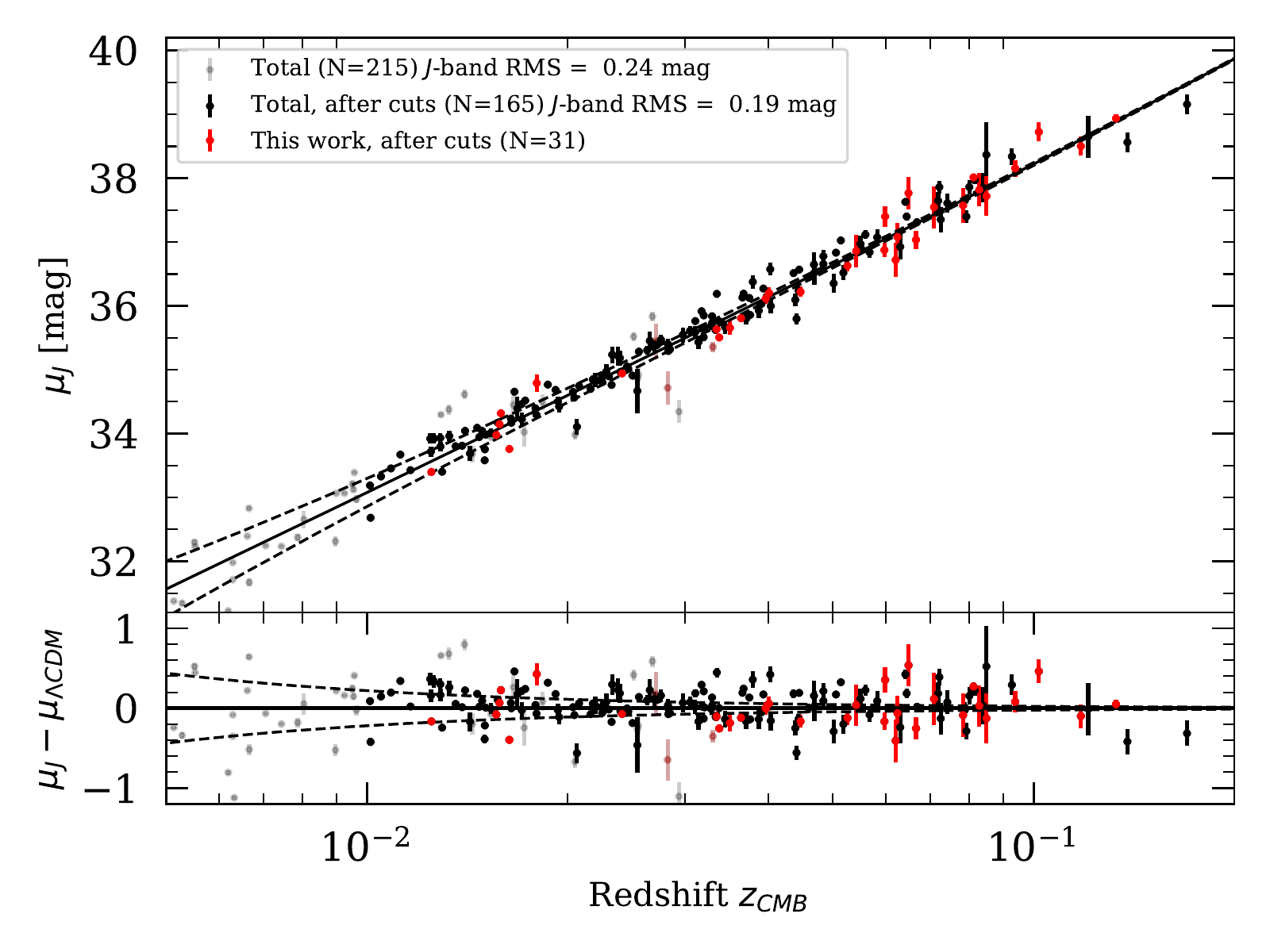}
    \includegraphics[angle=0,width=1.0\columnwidth]{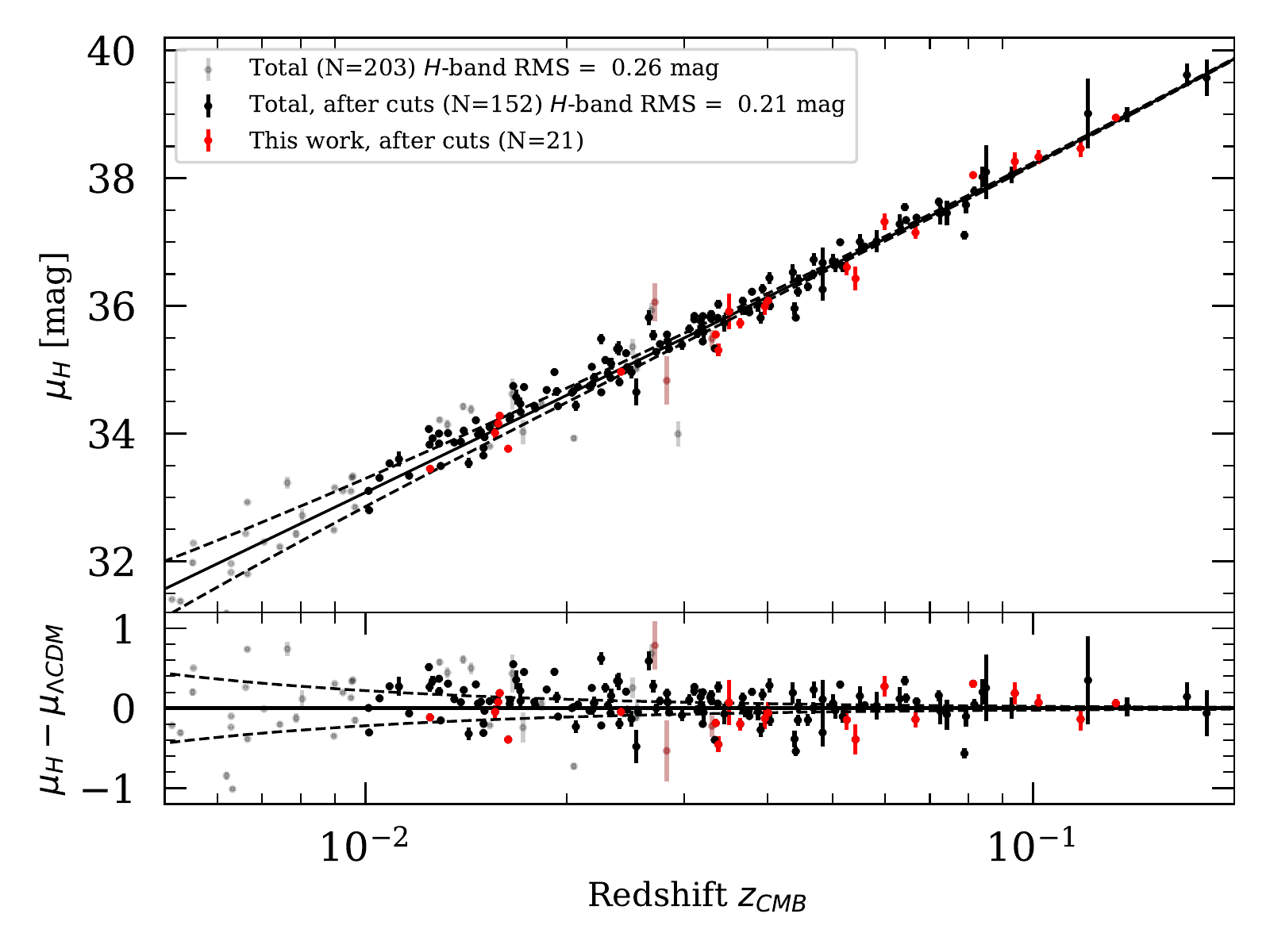}
    \caption{$J$ and $H$-band Hubble diagram and Hubble residuals for the SNe surving our cuts (165 in $J$, 152 in $H$). Red symbols show the SNe presented in this paper, and black symbols the SNe from the literature. Dashed lines indicate the scatter expected due to peculiar velocities $v_{\rm pec} \sim \pm 300$ \kms. 
    }
    \label{fig:hubble_JH}
\end{figure}

Figure \ref{fig:hubble_JH} shows the resulting $J$ and $H$-band Hubble diagrams for our optical+NIR SNe Ia compilation, including 40 of SNe from our sample presented.  iPTF14apg and iPTF14atg, are not included here, as we do not include spectroscopically peculiar SNe Ia (03fg, 06bt, 02es-like nor Iax SNe) in the analysis. Furthermore we apply a set of cuts on the redshift, stretch and color excess distribution on our sample, such that we include only SNe with $z_{CMB}>0.01$, $s_{BV}>0.5$, $E(B-V)_{\rm host}<0.5$ mag, and $E(B-V)_{\rm MW}<0.2$ mag (corresponding to typical sample cuts used in other cosmological analyses,  e.g. using SALT2 parameters $-0.3<c<0.3$ and $-3<x_1 <3$). The solid lines show the best-fit Hubble lines and the dashed lines indicate the scatter expected due to peculiar velocities $v_\mathrm{pec}=300$ km/s.

The RMS scatter in the Hubble residuals for the combined sample, after the cuts, is $\sigma_{HR, J}$=0.19 mag (165 SNe) and $\sigma_{HR, H}$=0.21 mag (152 SNe), for $J$ and $H$ respectively. The scatter in $J$ and $H$ does not decrease significantly when using individual best-fit $R_V$ instead of a global $R_V$.

We note an offset of $0.20\pm0.05$ mag when comparing the $Y$-band peak magnitudes to the CSP-I sample (also seen when comparing individual $Y$-band light curves of SNe observed simultaneously by RATIR and CSP-II, private communication). We thus add 0.20 mag to the $Y$-band magnitudes listed in Table~\ref{tab:peakmags} for the Hubble diagram analysis.

\subsection{Correlations with host galaxy stellar mass}
Having SN host galaxy stellar masses determined in Sect.~\ref{sec:hostmasses} and color and stretch corrected distances from Sect.~\ref{sec:hubblediagram}, we can begin to look for correlations.

\begin{figure}
    \centering
    \includegraphics[angle=0,width=0.5\textwidth]{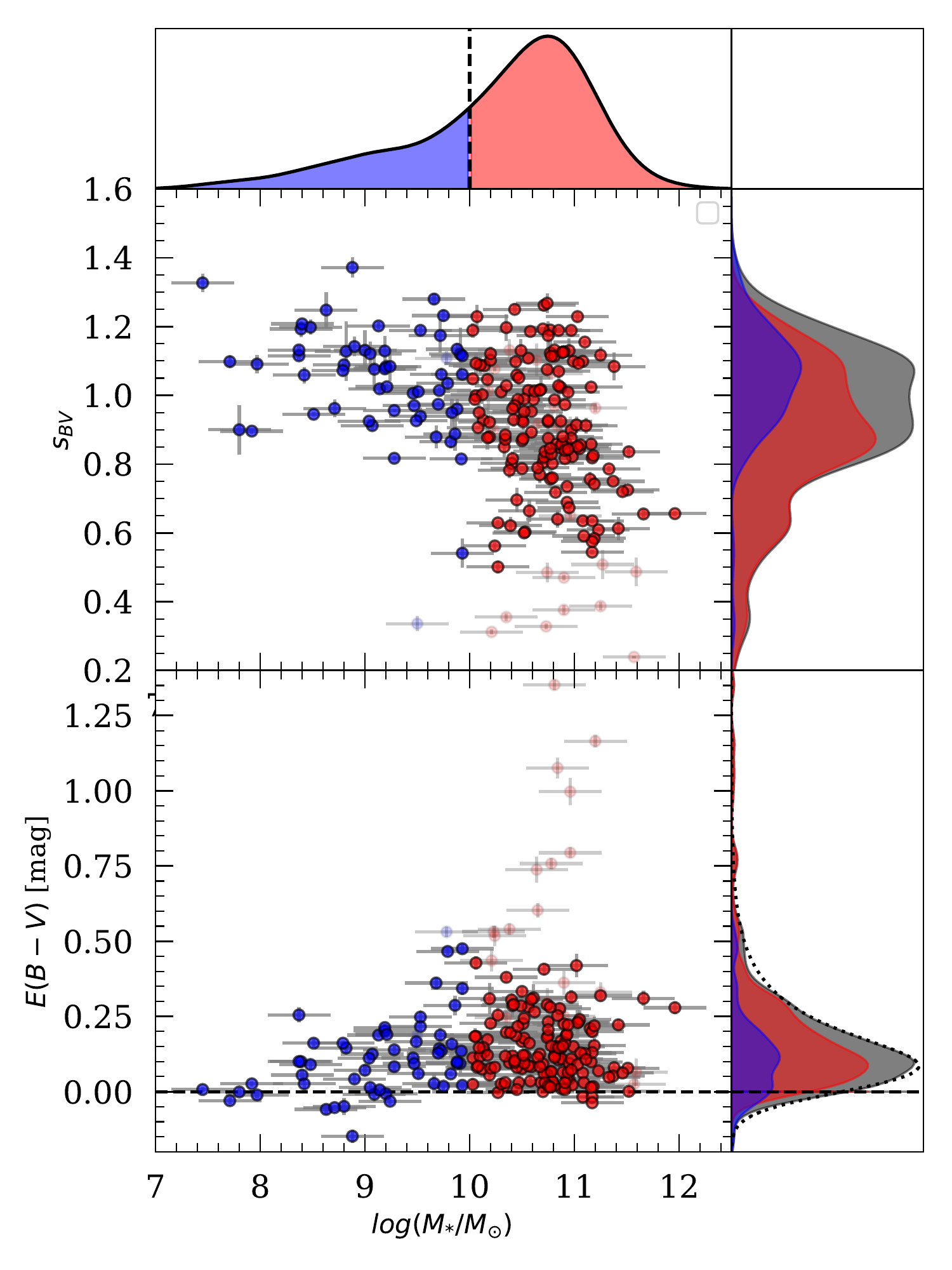}
    \caption{Upper panel: Distribution of stretch versus host galaxy mass. Low-mass galaxies (blue symbols) preferentially host slow-declining ($s>0.8$) SNe, while high-mass galaxies (red symbols) also host fast-declining SNe.
    Bottom panel: Distribution of fitted $E(B-V)_{\rm host}$ versus host galaxy mass. SNe in low-mass hosts typically have little reddening ($E(B-V) \lsim 0.25$ mag.), while there is a tail of highly extincted SNe occurring in high-mass galaxies.
    }
    \label{fig:st_EBV_lmass}
\end{figure}

In  Figure~\ref{fig:st_EBV_lmass} we show how our derived color stretch and color excess correlate with host stellar mass. Similar to conclusions reached in previous studies \citep{sullivan2011,childress2013}, we find that low-mass galaxies tend to host SNe with higher stretch ($s_{BV} > 0.8$) with moderate extinction ($E(B-V)_\mathrm{host} \lsim 0.25$ mag), while high-mass galaxies also host highly reddened SNe and fast-declining SNe.  

Following \citet{stanishev2018} and references therein, we fit the probability density function (PDF) of the computed color excesses for the entire sample, using an exponentially modified Gaussian distribution with a mean $c_0$ and standard deviation $\sigma_{c}$ and exponent relaxation parameter $\tau$.
We find values $c_0$=0.02 mag and $\sigma_c$=0.06 mag and $\tau$=0.14. We interpret the Gaussian component as a residual scatter due to intrinsic color variations.

Previous analyses \citep[e.g.][]{sullivan2011,betoule2014} typically split the sample at $M_{\rm split} = 10^{10}$ \Msun, which seems to be an "astrophysically reasonable" choice given the fairly distinct difference between the stretch and color excess distributions below and above \lmass=10.0. 
Other analyses have chosen a "statistically motivated" mass split location, either at the median stellar mass (\lmass$\sim$10.5) of their respective sample or based on some information criterion that maximizes the likelihood \citep{uddin2020,ponder2020,thorp2021}. We choose to split our sample at $M_\mathrm{split} = 10^{10}$ \Msun, as our fiducial case. Despite adding more SNe in low-mass galaxies from our sub-sample and e.g. the \citet{baronenugent2012} sub-sample, the distribution of host stellar mass for our sample is still skewed towards higher \lmass. For the combined sample, the median \lmass=10.50. 

\begin{figure*}
    \centering
    \includegraphics[angle=0,width=0.45\textwidth]{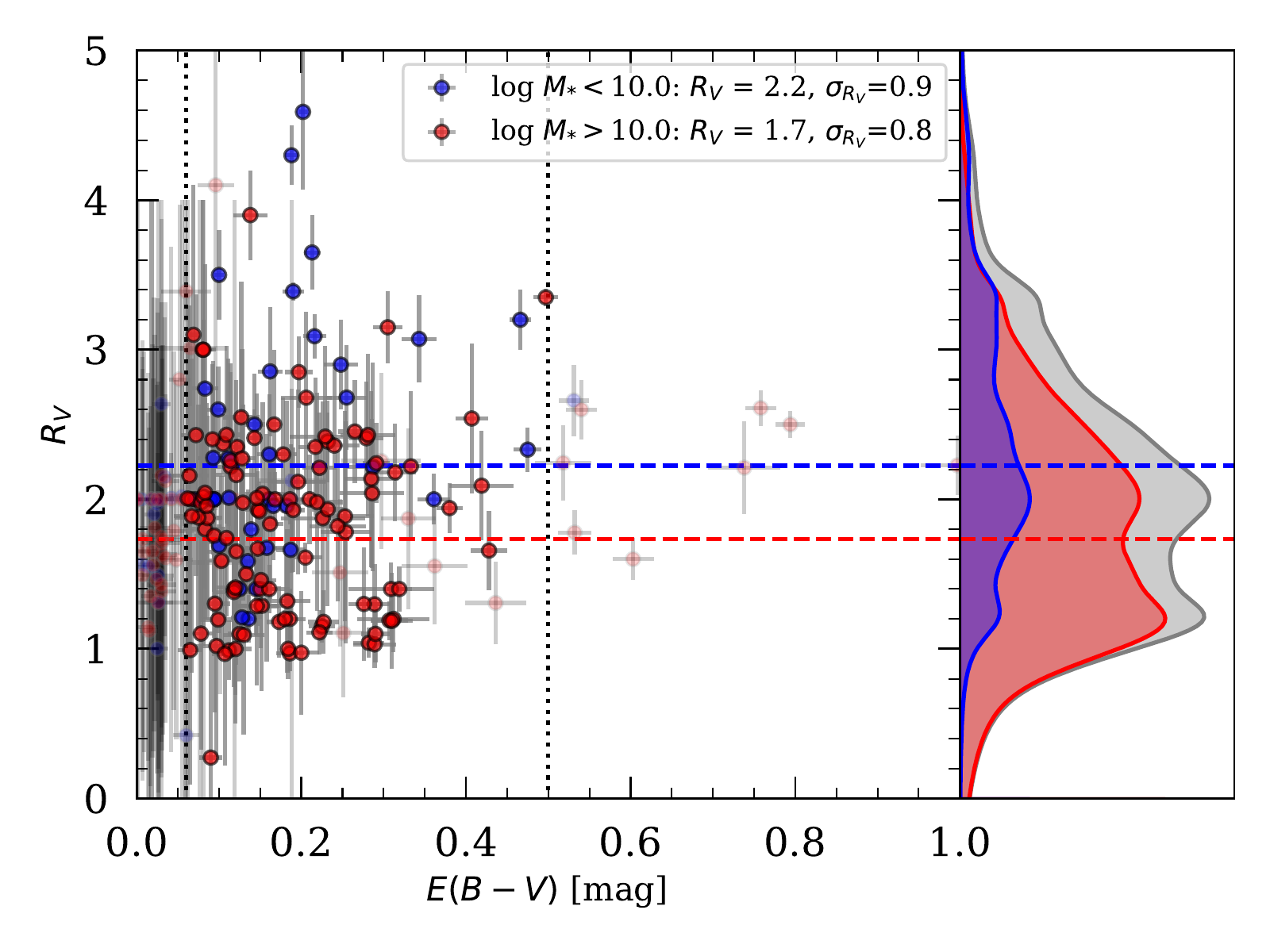}
\includegraphics[angle=0,width=0.45\textwidth]
{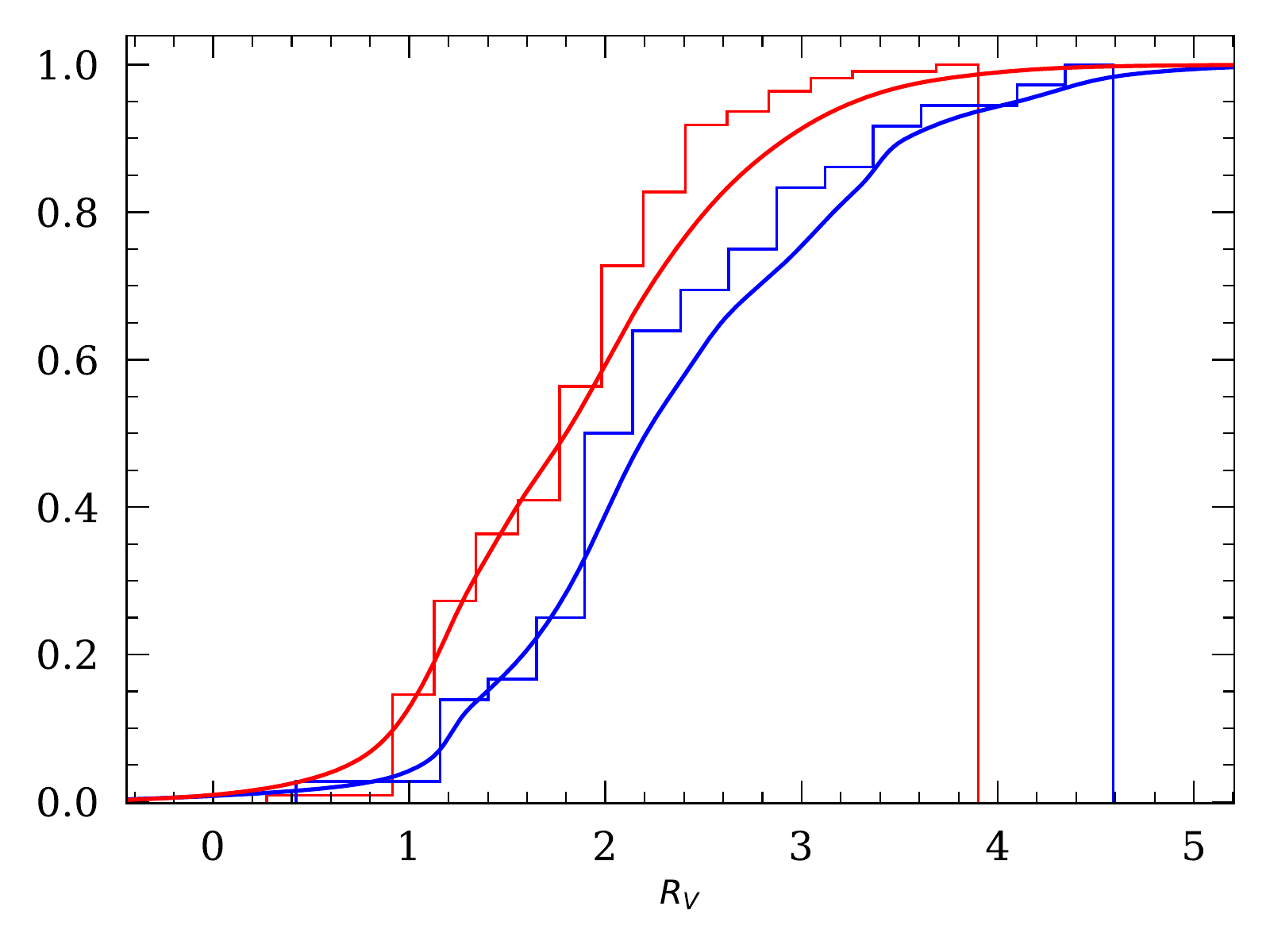}
    \caption{Left panel shows the distribution of best-fit $R_V$ and $E(B-V)_\mathrm{host}$ for the SNe included in the analysis. The color indicates if the SN occurred in a low- (blue) or high-mass host (red). The middle panel shows the coadded probability distribution of the best-fit $R_V$ values. The right panel shows the cumulative $R_V$ distribution for the low- and high-mass galaxies in blue and red, respectively.}
    \label{fig:rvdistribution}
\end{figure*}

If we look at the observed distribution of best-fit $R_V$ values (Fig.~\ref{fig:rvdistribution}), we find a weighted average
$R_V = 2.2$ ($\sigma_{R_V}= 0.9$) for \lmass$<10.0$ and $R_V = 1.7$ ($\sigma_{R_V} = 0.8$) for \lmass$>10.0$ host galaxies. The weighted average value of $R_V$ for the whole sample is $R_V = 1.9$ ($\sigma_{R_V} = 0.9$).

Here, we are not including $R_V$ estimates for SNe with color excesses close to the level of intrinsic color scatter $E(B-V)_\mathrm{host} < \sigma_c \sim 0.06 $ mag (where we typically find artificially low $R_V$, albeit with large error-bars) nor for highly extinguished SNe  $E(B-V)_\mathrm{host}>0.5$ mag (which are well fit by $R_V$ values ranging from 1.1 to 2.7, but the distribution is likely to be observationally biased towards finding SNe with low $R_V$).

In order to test the hypothesis that the distributions of $R_V$ in the low and high stellar mass bins are statistically compatible from being drawn from the same underlying distribution we perform a Kolmogorov-Smirnov (K-S) test. The K-S test yields a $p$-value of only 0.015, hence suggesting that the distributions are significantly different, with more than 95\% confidence level. 

Even though the weighted mean values are statistically consistent with the global mean $R_V$ value, we stress that the wide, non-gaussian, probability distributions are different.

\begin{figure*}
    \includegraphics[angle=0,width=\columnwidth]{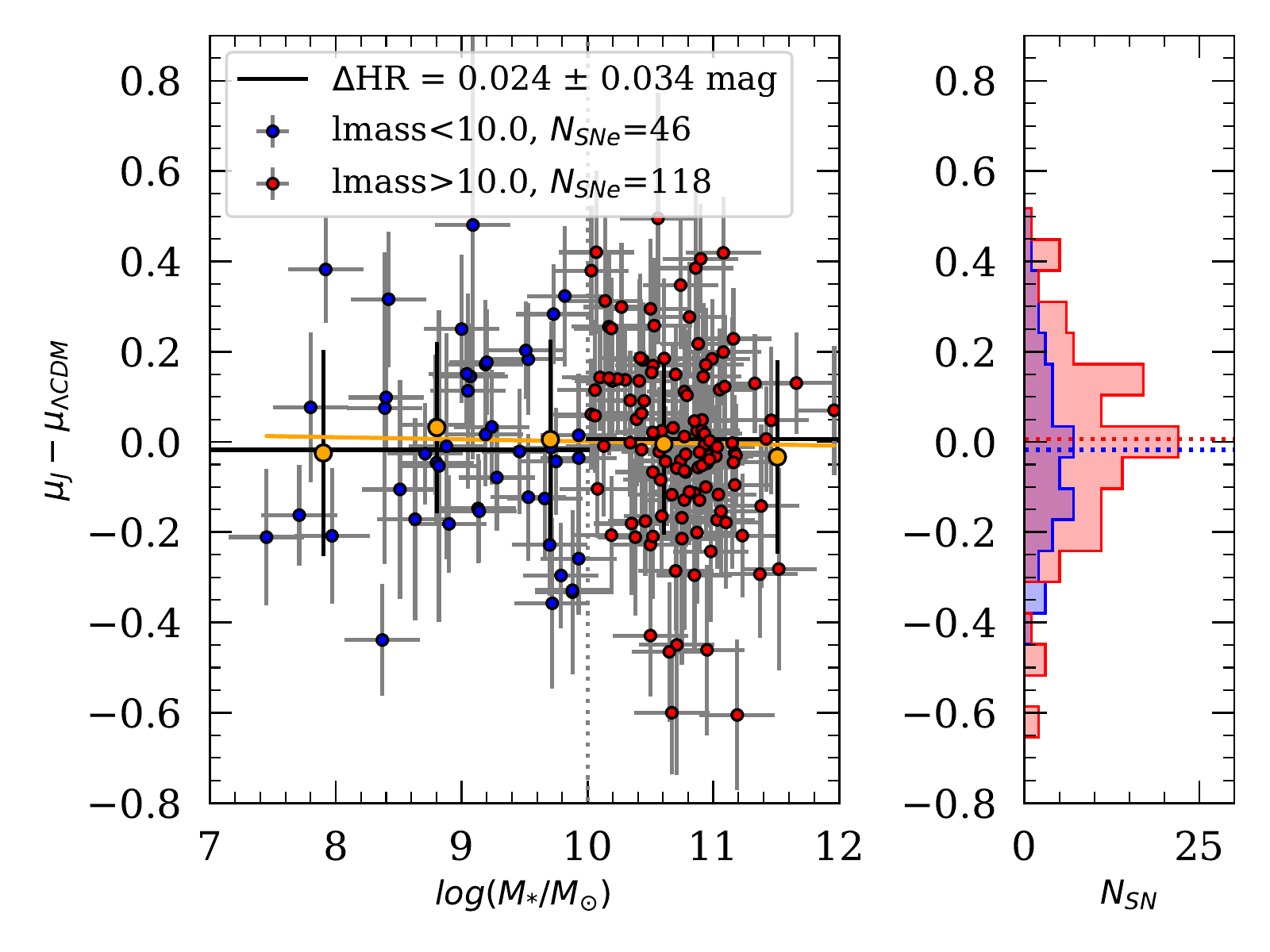}
    \includegraphics[angle=0,width=\columnwidth]{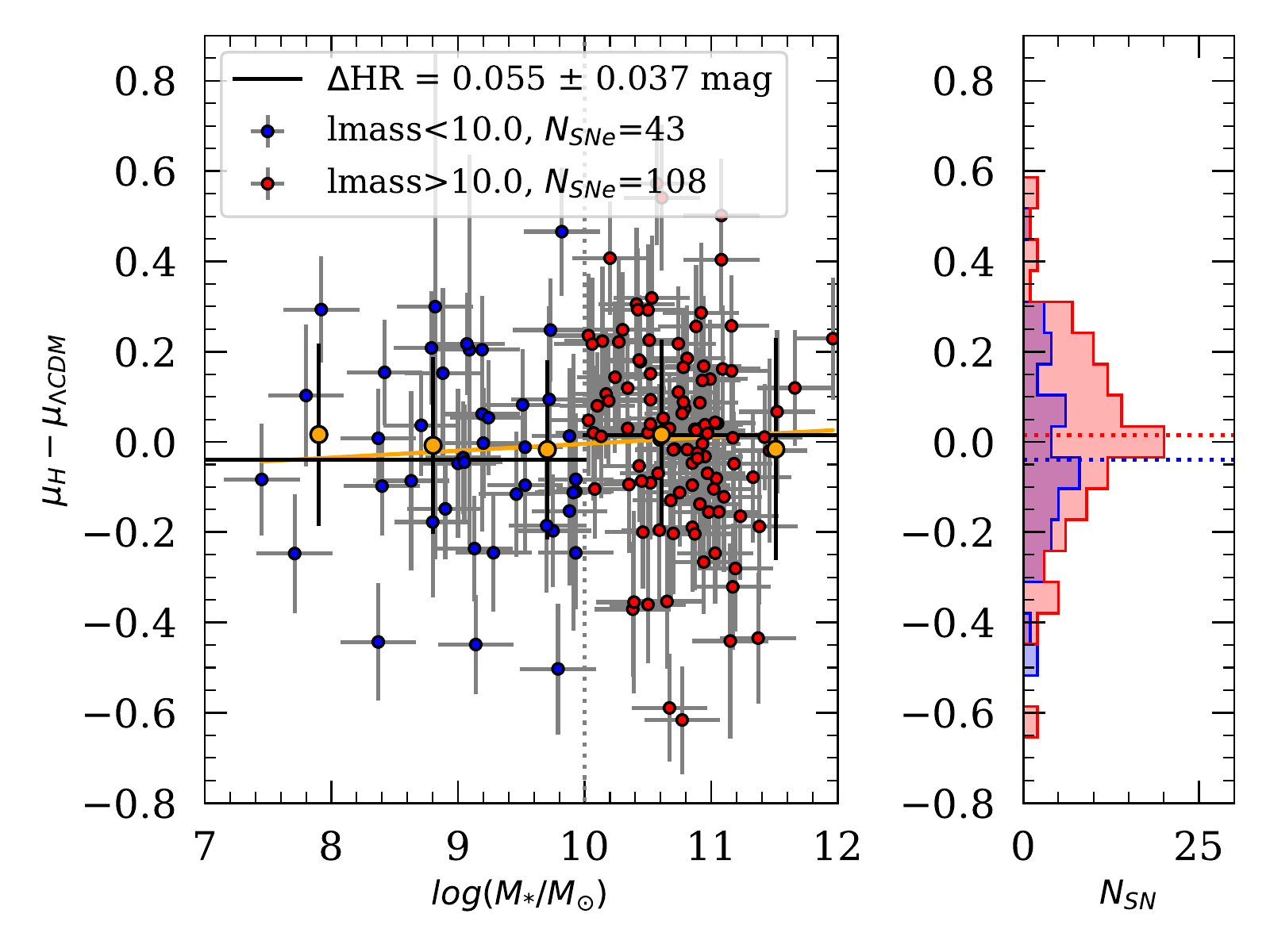}
    \caption{$J$ and $H$-band Hubble residuals versus host galaxy stellar mass from fitting optical and NIR lightcurves with the \texttt{color\_model} (each SN corrected with best-fit $E(B-V)_\mathrm{host}$ and $R_V$). Orange symbols show the binned mean and standard deviation of the Hubble residuals in five mass bins, while the orange line is the fitted slope ($-0.01 \pm 0.02$ mag/dex in $J$, $0.0 \pm 0.02$ mag/dex in $H$).}
    \label{fig:HR_JH}
\end{figure*}

\begin{figure*}
    \centering
    \includegraphics[angle=0,width=1.0\columnwidth]{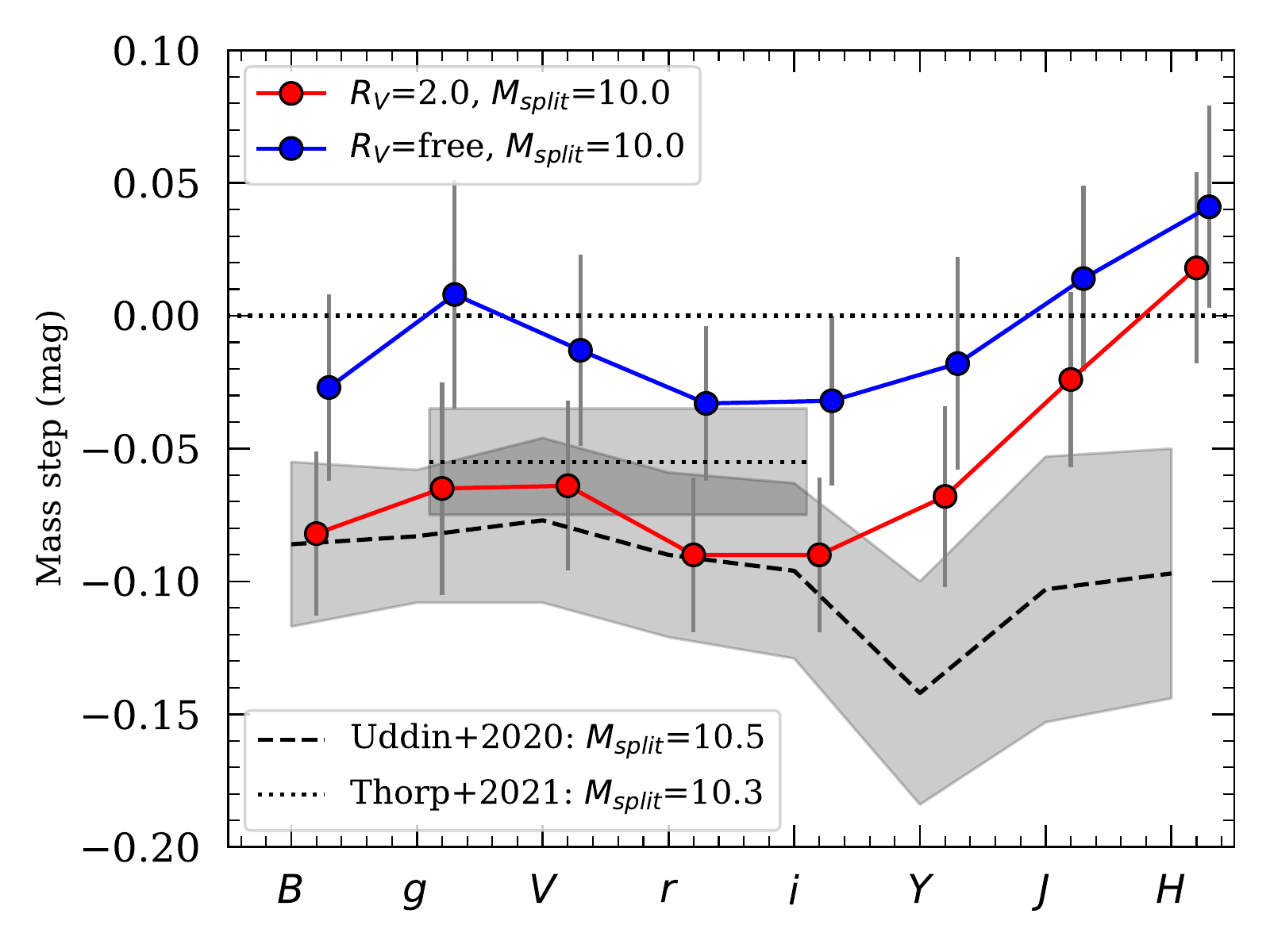}\includegraphics[angle=0,width=1.0\columnwidth]{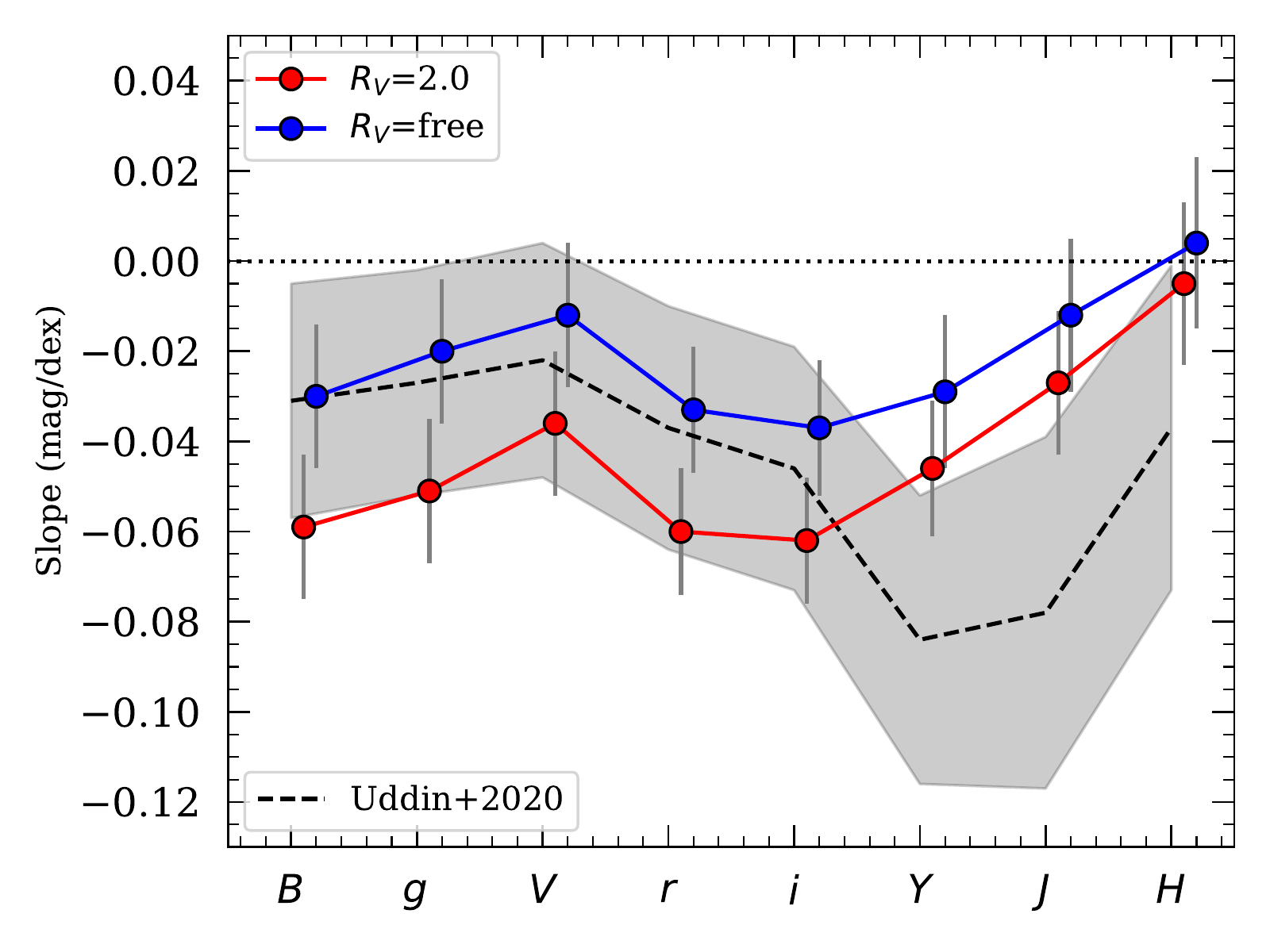}
    \caption{Size of the mass-step (left panel) or mass-slope (right panel) in different filters ($BgVriYJH$). Blue symbols show the resulting mass-steps/slopes if we use the best-fit $R_V$ for each individual SN, and red symbols if we use a global $R_V=2.0$.
    Gray dashed lines show the mass-steps from \citet{uddin2020} and (their Table 2) for the same cuts on $E(B-V)_\mathrm{host}<0.5$ mag and $s_{BV}>0.5$, but for $\log M_{\rm split}=10.5$ and including SNe at $z<0.01$. The dotted line shows the mass-step in \citet{thorp2021}.
    }
    \label{fig:mass_steps}
\end{figure*}

\begin{figure*}
    \centering
    \includegraphics[angle=0,width=0.4\textwidth]{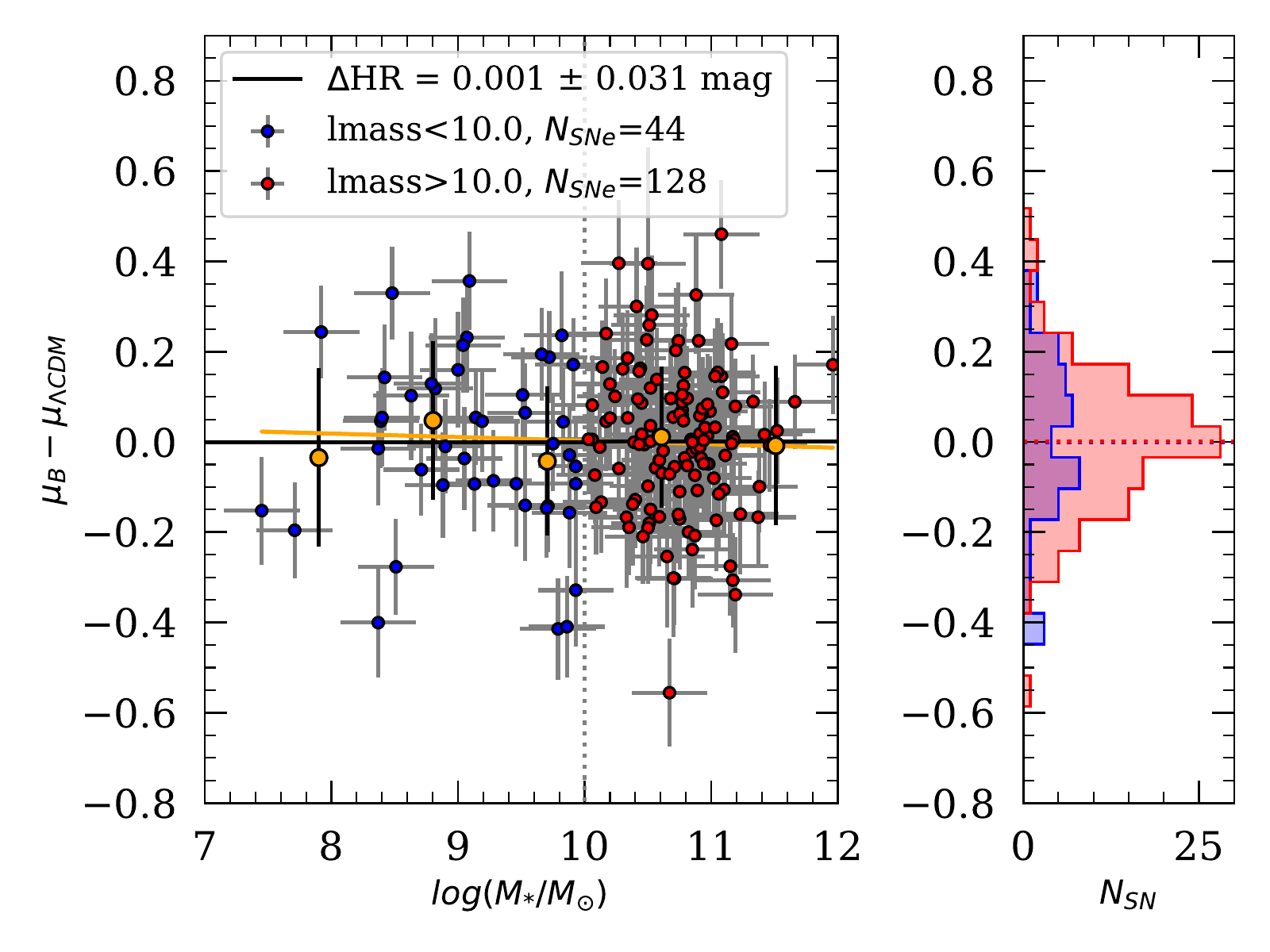}\includegraphics[angle=0,width=0.4\textwidth]{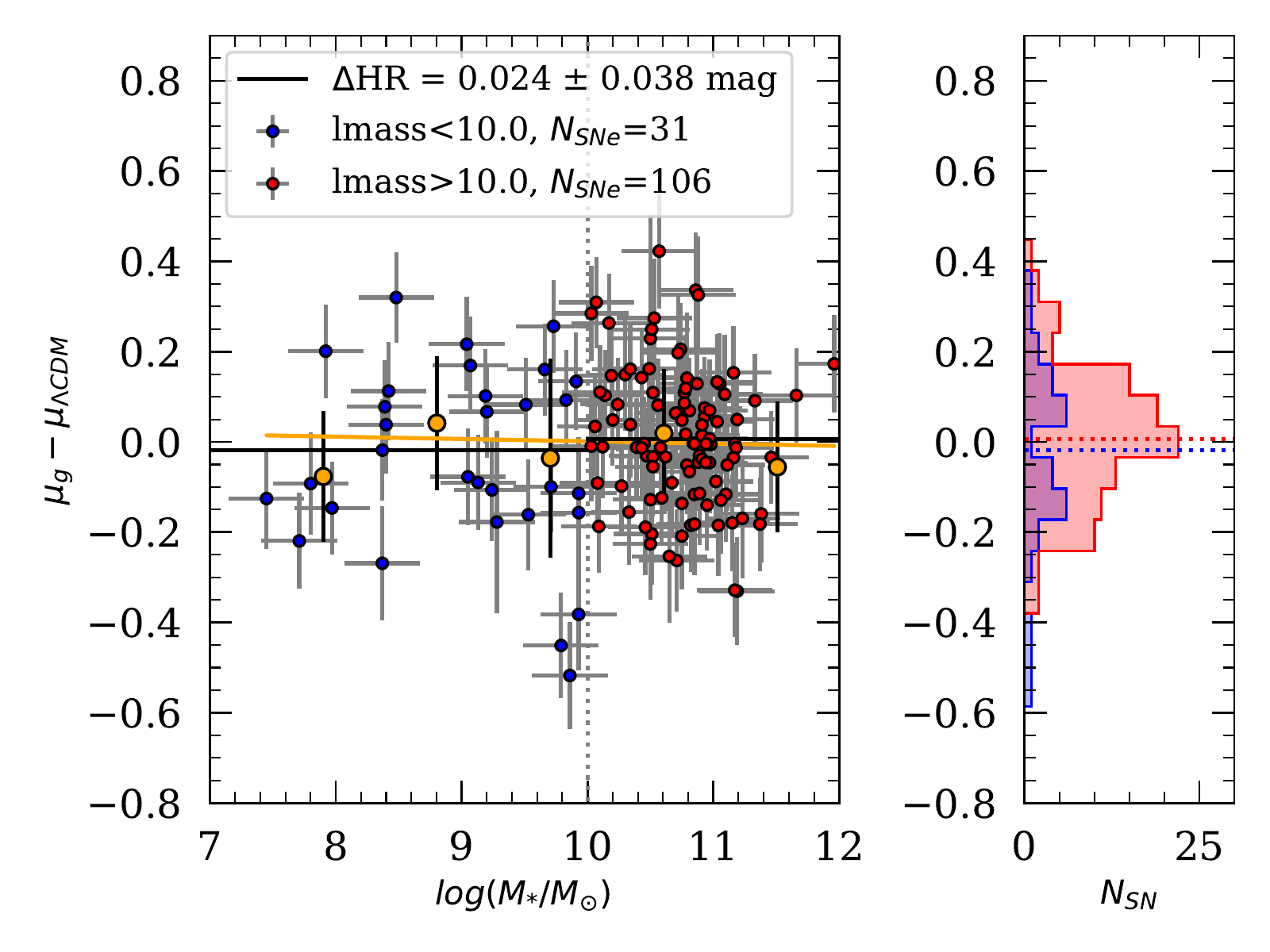}
    \includegraphics[angle=0,width=0.4\textwidth]{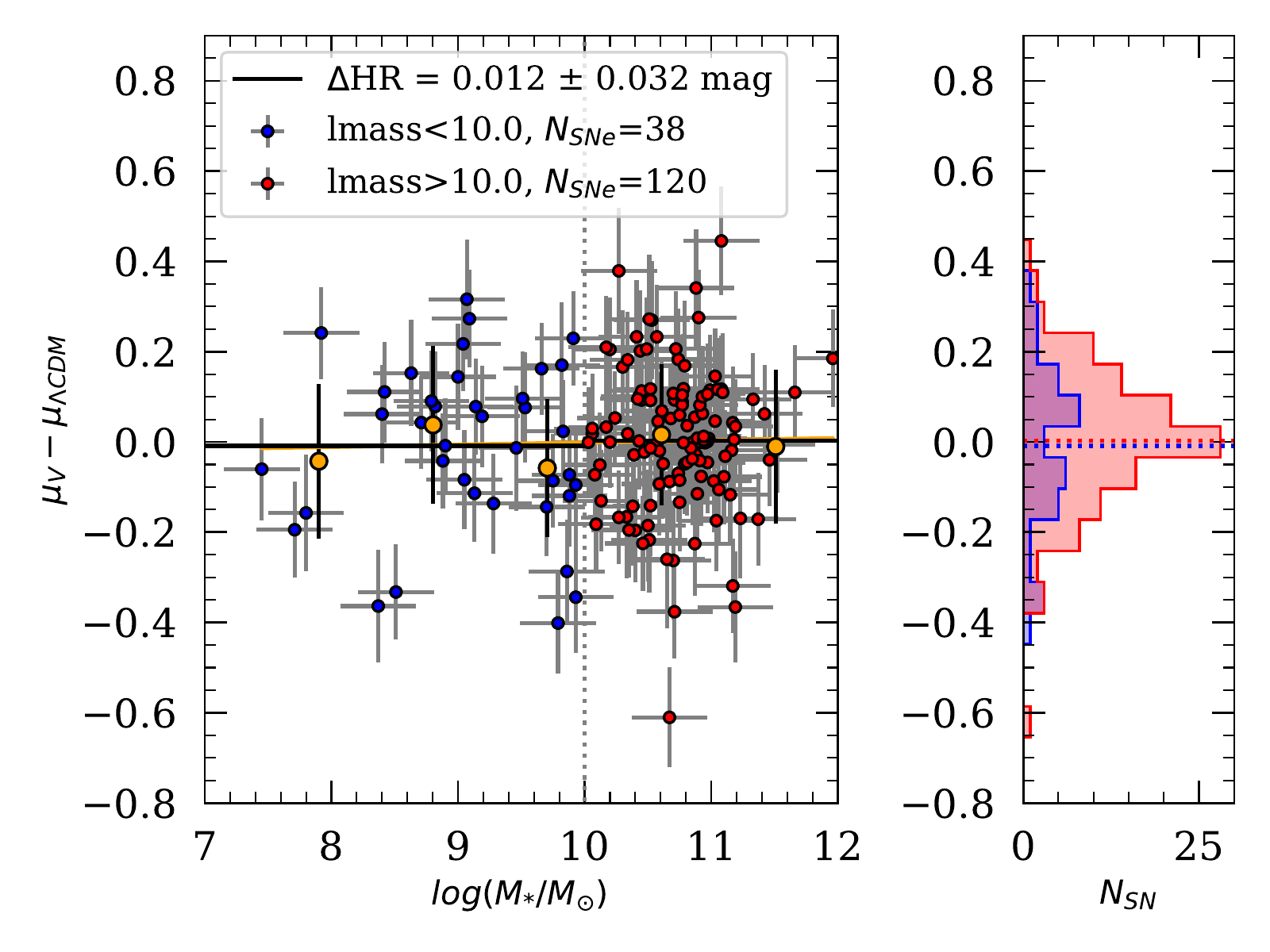}\includegraphics[angle=0,width=0.4\textwidth]{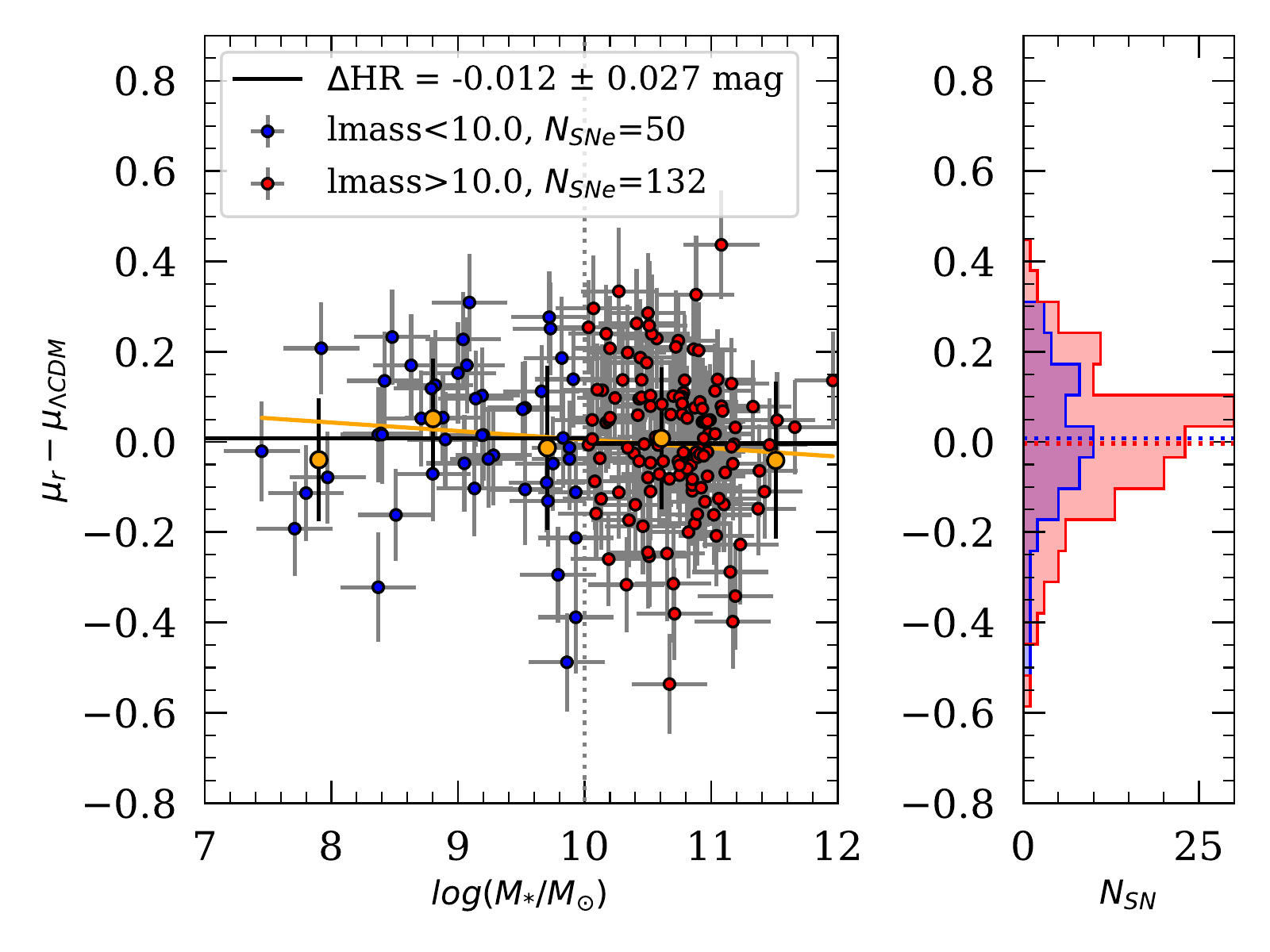}
    \includegraphics[angle=0,width=0.4\textwidth]{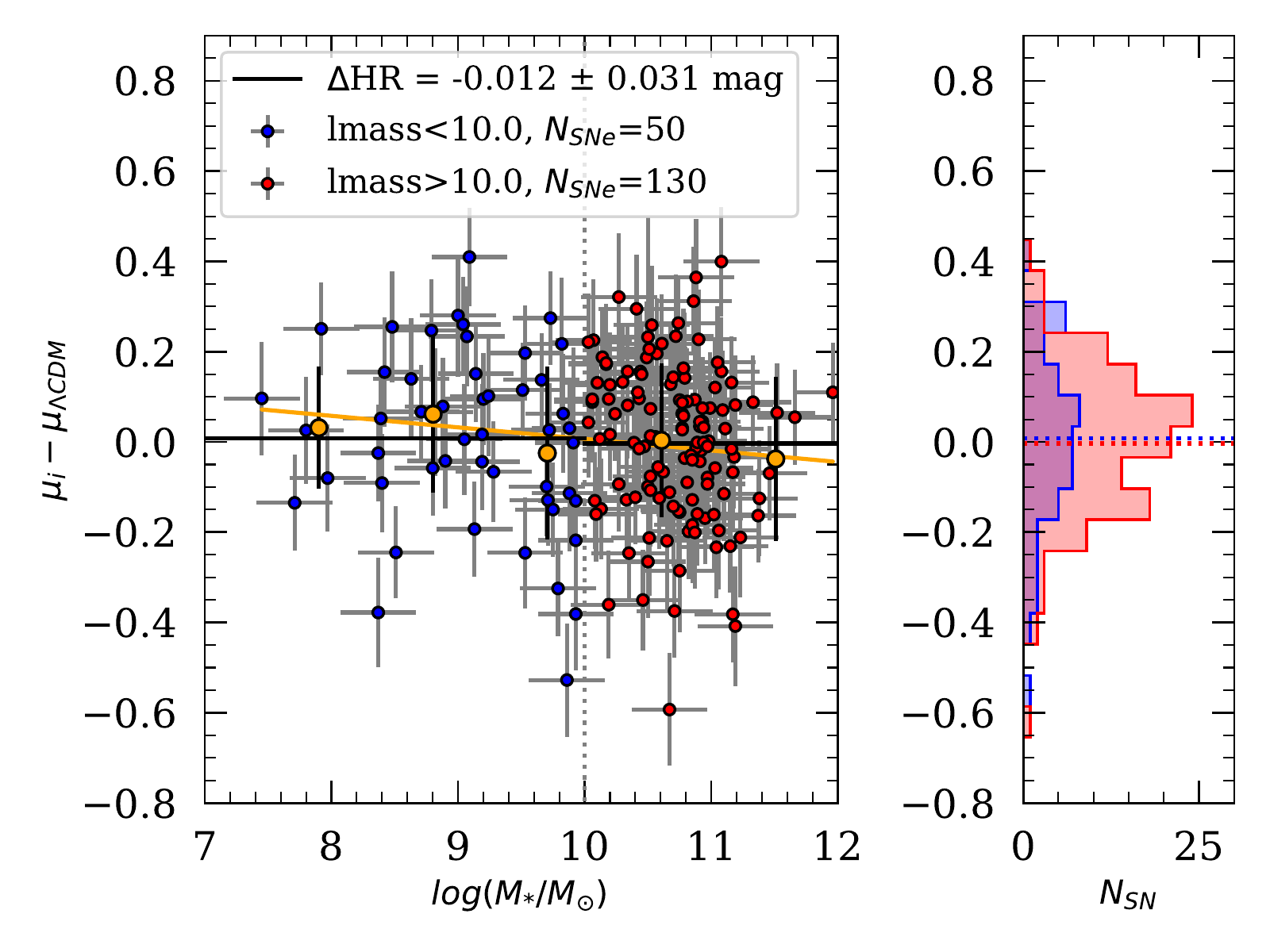}\includegraphics[angle=0,width=0.4\textwidth]{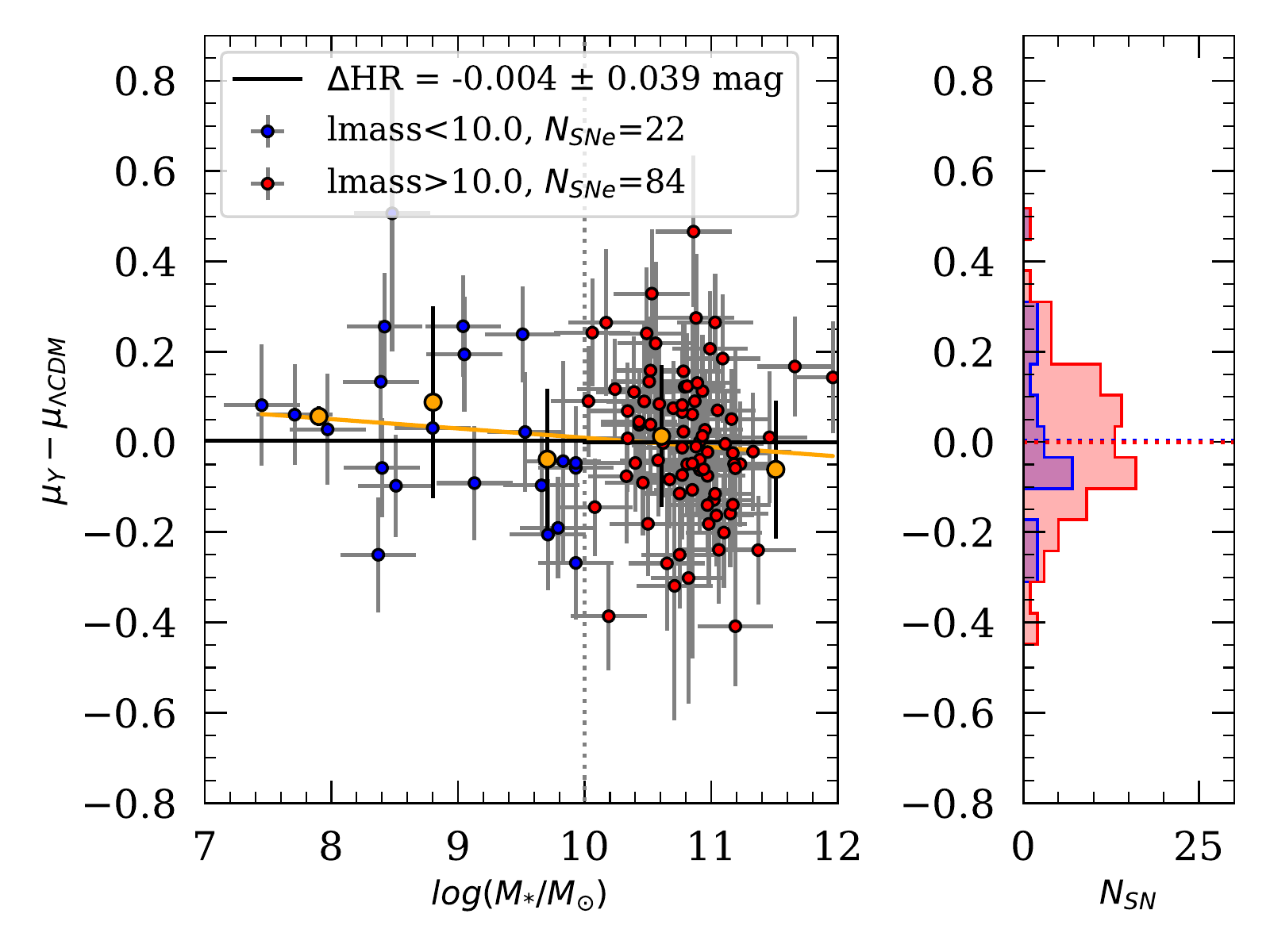}
    \includegraphics[angle=0,width=0.4\textwidth]{figures/HR_J.pdf}\includegraphics[angle=0,width=0.4\textwidth]{figures/HR_H.pdf}
    \caption{Hubble residuals versus host galaxy stellar mass from fitting from optical ($BgVri$) and NIR ($YJH$) lightcurves with the \texttt{SNooPy color\_model} (i.e. each SN corrected with best-fit $E(B-V)_\mathrm{host}$ and $R_V$). Orange symbols show the binned mean and standard deviation of the Hubble residuals in five mass bins, while the orange line is the fitted slope.
    }
    \label{fig:HR_all}
\end{figure*}

For the low- and high-mass bins, we compute a weighted mean and standard deviation of the Hubble residuals for $BVgriYJH$ filters (horizontal black lines in Fig.~\ref{fig:HR_JH} and \ref{fig:HR_all}). For each filter $X$, we will refer to any Hubble residual offset between the two bins as a "mass-step", $\Delta_{HR}(X)$. To further investigate the behaviour of the Hubble residuals, we divide the sample into five mass bins (orange symbols in Fig.~\ref{fig:HR_JH}), to see if there are any additional effects towards the edges of the host mass distribution. Following \citet{uddin2020}, we also fit a slope to the Hubble residuals as a function of host mass (yellow lines in Fig.~\ref{fig:HR_JH} and \ref{fig:HR_all}) using Orthogonal Distance Regression (ODR). 

We find that using a global value $R_V=2.0$ (close to the average $R_V$ for the entire sample) for all SNe in low- and high-mass host galaxies, we reproduce a significant ($\sim 2\sigma$) mass-step in optical $BVgri$-band Hubble residuals $\Delta_{HR} \sim -0.07 \pm 0.03$ mag, while for NIR $JH$-bands there is no significant mass step ($\Delta_{HR}(J) = -0.021 \pm 0.033$ mag and $\Delta_{HR}(H) = 0.020 \pm 0.036$ mag), shown as red symbols  in Fig.~\ref{fig:mass_steps} (left panel). A similar trend is seen when fitting a slope to the Hubble residuals as a function of host mass. For optical $BVgri$-bands we find a slopes of $\sim 0.06 \pm 0.02$ mag/dex, while in the NIR the slopes are smaller ($-0.027 \pm 0.016$ mag/dex and $-0.005 \pm 0.018$ mag/dex in $J$ and $H$, respectively) shown as red symbols in right panel of Fig.~\ref{fig:mass_steps}. 

When correcting each SN individually by their best-fit $R_{V} \times E(B-V)_{host}$ we see no significant mass-step or mass-slope in the Hubble residuals, across the optical and NIR bands (blue symbols in Fig.~\ref{fig:mass_steps}). 

This result seems valid when changing the cuts on $z$, $s_{BV}$ and $E(B-V)_\mathrm{host}$, and perhaps more importantly the choice of $M_\mathrm{split}$. Choosing $M_{\rm split} = 10^{10.5}$ \Msun (the median stellar host galaxy mass of our SNe compilation), we do see a (non-significant) mass-step across optical and NIR $\Delta_{HR} \sim -0.04 \pm 0.03$ mag. 
We find that our results are in line with with \citet{broutscolnic2021}, who modelled host galaxy reddening as separate Gaussian distributions for galaxies below and above \lmass=10. They found that SNe in low-mass hosts, the average $\langle R_V \rangle = 2.75 \pm 0.35$, whereas for SNe in high-mass hosts, $\langle R_V \rangle = 1.5 \pm 0.25$, with both sub-samples having a wide distributions $\sigma_{R_V} = 1.3$. This is in fair agreement with \citet{Salim2018}, who find that on average, dusty, high-mass quiescent galaxies have lower $R_V$ values ($\langle{R_V}\rangle=2.61$), whereas low-mass star forming galaxies tend to have higher values for $R_V$ ($\langle{R_V}\rangle=3.15$). 

\citet{uddin2020} found nominal evidence for a consistent mass-step in both the optical and NIR using the CSP-I sample ($\Delta_{HR,J} = -0.103 \pm 0.050$ mag, and $\Delta_{HR,H} = -0.097 \pm 0.047$ mag) using similar cuts on the sample, although including SNe with $z<0.01$ and having the mass-step located at \lmass=10.5 (shown as gray dashed lines in Fig.~\ref{fig:mass_steps}).
We can not fully reproduce the NIR mass-step reported by \citet{uddin2020}, even if we use their host masses and the same best-fit extinction. We note that they do use updated Phillips-relations, correcting for stretch using more flexible spline functions calibrated using unpublished data from CSP-II (C. Burns and S. Uddin, private communication), but we do not expect this to be result in the observed differences in our plots.

\citet{ponder2020} also report a $H$-band mass-step $\Delta_{HR,H} = -0.18 \pm 0.05$ mag (mass-step located at \lmass=10.44) using a compilation of 99 SNe from the literature. However, after removing two outliers, the step reduces to $\Delta_{HR,H} = -0.10 \pm 0.04$ mag. It is unclear if their results are due to the lack of NIR stretch corrections and/or color-corrections. 

Recently, \citet{thorp2021} analyzed optical ($griz$) lightcurves of 157 nearby SNe Ia (0.015 $< z <$ 0.08) from the Foundation DR1 dataset using the BayesSN lightcurve fitter \citep{mandel2011,mandel2020}. When splitting their sample at \lmass=10, they find $R_V = 2.84 \pm 0.31$ in low-mass hosts, and $R_V = 2.58 \pm 0.23$ in high-mass hosts. They conclude that these values are consistent with the global value of $R_V = 2.61 \pm 0.21$, estimated for the full sample, and can not be an explanation of the mass step. After corrections, their resulting mass-step is $\Delta_{HR}=0.050 \pm 0.022$ mag (shown as gray dotted line in left panel of Fig.~\ref{fig:mass_steps}).

\section{Discussion}
\label{sec:discussion}
The relation between SN~Ia luminosity and host galaxy properties is of great interest for SN~Ia progenitor studies as well as for cosmology, as a third empirical correction \citep{sullivan2010,2010ApJ...722..566L,2010ApJ...715..743K}. While the correlation has been seen nearly ubiquitously across different samples \citep[see e.g.][]{Brout18-SYS,smith2020}, there is a significant debate about the physical origin of this relation. There has been speculation that the mass-step may be driven by the age of the stellar population, metallicity or star formation rate \citep{sullivan2010,2011ApJ...740...92G,2011ApJ...743..172D,2013ApJ...764..191H,Childress2014,rigault2020,Rose2020}. However, it is possible that this correlation arises due to different SN~Ia environments in different host galaxy types. 

While most studies pertaining to SN luminosity and host galaxy correlations in the optical, recently there have been reports of the possible detection of the mass-step in the NIR wavebands \citep{ponder2020,uddin2020}.  
In this study, we exploit the multi-wavelength, well-sampled lightcurves of the SNe in our sample and compute the mass-step/slope after stretch and color corrections, fitting the $R_V$ parameter individually for each SN. We find that when fitting the $R_V$ value for each SN, we see a mass step consistent with zero in all filters from $B$ to $H-$band (see Figure~\ref{fig:mass_steps}). However, when fixing the $R_V$ value to the sample average, as is done in previous studies, we find that there is a mass step of $\sim$ 0.07 - 0.1 mag in the optical ($BVgri$) while no significant step is seen in the NIR ($YJH$).

Since the free $R_V$ case yields mass steps that are consistent with zero, it is likely that the origin of the mass step is due to variations of dust properties in the interstellar medium of the host galaxies. However, more detailed studies would be required to rule out intrinsic effects. In fact, there are indications that there are two SN populations, having different SN ejecta velocities and different intrinsic colors, which also trace the host galaxy stellar mass \citep[see e.g.][]{polin2019,siebert2020,pan2020}.
\citet{childress2013} and \citet{Gonzalez-Gaitan2020} simulate the effect of having separate color-luminosity corrections for low- and high-mass galaxies. They find that multiple free color-luminosity slope parameters may explain away the mass-step, suggesting that the origin of mass-step is a difference in intrinsic color-luminosity relation ($\beta_{int} c_{int}$) of two SN populations found in galaxies with different masses as opposed to different dust properties.

\section{Conclusion}
\label{sec:conclusion}
Many studies in the literature have suggested the need for an additional standardization parameter for SNe Ia, beyond the lightcurve shape and color. In particular, firm evidence has been put forward for a correlation between residuals in the Hubble diagram at optical wavelengths, and the host galaxy stellar mass \citep{sullivan2010,2010ApJ...722..566L,2010ApJ...715..743K,2011ApJ...740...92G,2011ApJ...743..172D,2013ApJ...764..191H,Childress2014,Rose2020}. 
The underlying cause of these correlations is not completely understood, with some suggestions that this could be due to correlation with age/metallicity of the underlying stellar population, however, there is also evidence pointing to this correlation arising from differences in dust properties of the SN hosts. 

Our work differs from previous studies in that we use a sample of SNe Ia, found in the untargeted PTF/iPTF surveys,  which adds a significant number of low-mass host galaxies. Furthermore, our data-set includes multi-wavelength follow-up observations, including near-IR, which allows us to infer the total-to-selective extinction parameter, $R_V$, for each SN individually. This is motivated by the findings in e.g. \citet{Amanullah2015} and \citet{burns2018}, suggesting that the wavelength dependence of dimming by host galaxy mass varies between SNe, making the use of a single value of $R_V$ questionable. 
Using a parameterized extinction relation by CCM, we fit for both the color-excess, $E(B-V)$, and $R_V$ using \texttt{SNooPy color\_model} fits of the multi-band data. In other words, our estimate of the extinction along the line-of-sight of each SN is not derived from the Hubble residuals. When examining the Hubble residuals, we do not find a significant correlation with stellar mass at optical or NIR wavelengths. 

If we, instead, assume a single value for $R_V$ to correct all SNe fitting only the color excess, we recover the "mass-step" in optical filters. In the NIR, we find no significant dependence on the stellar mass, independently of how $R_V$ is measured, i.e., individually or globally. This is consistent with the interpretation made by \citep{broutscolnic2021}, that the mass-step is likely caused by differences in dust properties of the low- and high-mass SN host galaxies.

\section{Acknowledgements}
We are grateful for valuable comments from Jakob Nordin and feedback from the CSP team (Chris Burns, Eric Hsiao, Mark Phillips and Shuvo Uddin). We thank M. L. Graham for helping to obtain follow-up observations with Las Cumbres Observatory in 2013.

A.G acknowledges support from the Swedish Research Council under Dnr VR 2016-03274 and 2020-03444. T.P. acknowledges the financial support from the Slovenian Research Agency (grants I0-0033, P1-0031, J1-8136 and Z1-1853).
Research by S.V. is supported by NSF grants AST-1813176 and AST-2008108.

This work makes use of observations from the Las Cumbres Observatory network. The LCO team was supported by NSF grants AST-1313484 and AST-1911225. The intermediate Palomar Transient Factory project is a scientific collaboration among the California Institute of Technology, Los Alamos National Laboratory, the University of Wisconsin, Milwaukee, the Oskar Klein Center, the Weizmann Institute of Science, the TANGO Program of the University System of Taiwan, and the Kavli Institute for the Physics and Mathematics of the Universe. LANL participation in iPTF is supported by the US Department of Energy as a part of the Laboratory Directed Research and Development program
\bibliography{nirpaper}{}
\bibliographystyle{aasjournal}

\begin{table*}
  \centering
  \caption{42 SNe~Ia from the iPTF survey and their associated host galaxies.
  }
  \label{tab:sne}
  \begin{tabular}{@{}lllllll@{}}
    \hline \hline
SN 				        & $z_{\rm helio.}$    & $z_{\rm CMB}$ & R.A.			& Decl.			& Host galaxy       & $\log \left(\frac{M_{*}}{M_{\odot}} \right)$ \\
\hline
iPTF13S 		        & 0.059$^{a}$   & 0.060     & 203.222074 	& +35.959372	& SDSS J133253.27+355733.4 & 7.97	\\ 
iPTF13ez 		        & 0.04363       & 0.04470   & 182.463737 	& +19.787693	& KUG 1207+200              & 10.19 \\
iPTF13ft	 	        & 0.03884	    & 0.03963   & 199.947067 	& +33.024961	& SDSS J131947.32+330131.8  & 8.80 \\
iPTF13abc (SN\,2013bh)	& 0.07436 	    & 0.07498   & 225.554527 	& +10.645905	& SDSS J150214.17+103843.6  & 10.33 \\ 
iPTF13ahk               & 0.02639	    & 0.02712   & 203.805002 	& +34.678903	& NGC 5233                  & 11.27 \\
iPTF13anh	 	        & 0.0615$^{b}$	& 0.0625    & 196.710215	& +15.575657	& SDSS J130650.44+153432.7  & 8.51	\\ 
iPTF13aro               & 0.08462		& 0.08497   & 236.884308 	& +23.023956	& SDSS J154732.26+230111.8 & 10.75	\\ 
iPTF13asv (SN\,2013cv)	& 0.0362$^{b}$ 	& 0.0364    & 245.679971	& +18.959717	& SDSS J162243.02+185733.8 & 7.71	\\ 
iPTF13ayw	 	        & 0.05385		& 0.05418   & 234.889650 	& +32.093954	& SDSS J153933.08+320538.3 & 11.15	\\ 
iPTF13azs (SN\,2013cx)	& 0.0338$^{b}$  & 0.03376   & 256.067046 	& +41.510353	& SDSS J170415.96+413036.8 & 9.79	\\ 
iPTF13bkw 		        & 0.06393		& 0.06491   & 200.489840 	& +11.735753	& SDSS J132157.57+114406.2 & 10.56	\\ 
iPTF13crp			    & 0.0630$^{b}$	& 0.0621    & 29.750591 	& +16.264187	& SDSS J015900.28+161551.5 & 10.71	\\ 
iPTF13daw		        & 0.07755		& 0.07680   & 40.880381 	& +1.984422	    & SDSS J024331.69+015908.4 & 10.82 \\ 
iPTF13ddg	 	        & 0.084$^{a}$   & 0.083     & 11.961798 	& +31.821517	& SDSS J004750.94+314922.5 & 9.71	\\ 
iPTF13dge		        & 0.015854		& 0.015805  & 75.896169 	& +1.571493	    & NGC 1762                  & 10.87	\\ 
iPTF13dkj			    & 0.03623 	    & 0.03503   & 347.211539 	& +20.069088	& CGCG 454-001              & 10.50	\\ 
iPTF13dkx		        & 0.0345$^{b}$  & 0.0335    & 20.221425 	& +3.339925	    & SDSS J012052.56+032023.0 & 9.13	\\ 
iPTF13duj (SN\,2013fw)	& 0.016952		& 0.015879  & 318.436571 	& +13.575875	& NGC 7042                  & 10.99	\\ 
iPTF13dym 		        & 0.04213 		& 0.04091   & 351.125804 	& +14.651100	& SDSS J232430.20+143903.5 & 9.93	\\ 
iPTF13dzm		        & 0.018193 		& 0.017219  & 17.824325 	& +33.112441	& NGC 0414              & 10.25	\\ 
iPTF13ebh               & 0.013269		& 0.012493  & 35.499900 	& +33.270479	& NGC 890             & 11.23	\\ 
iPTF13efe		 	    & 0.070$^{a}$   & 0.071     & 130.913761 	& +16.177023	& SDSS J084339.26+161037.5 & 8.39	\\ 
iPTF14yw (SN\,2014aa)   & 0.016882      & 0.017972  & 176.264696 	& +19.973620	& NGC 3861              & 10.86	\\ 
iPTF14yy			    & 0.04311		& 0.04423   & 186.538205	& +9.978942	    & SDSS J122608.78+095847.1 & 10.40	\\ 
iPTF14aje		 	    & 0.02769		& 0.02825   & 231.300298 	& -1.814299	    & UGC 9839                  & 10.96	\\ 
iPTF14ale		 	    & 0.093226 		& 0.093835  & 219.587725 	& +27.334341	& SDSS J143822.02+272010.6 & 11.36	\\
iPTF14apg	 	        & 0.08717       & 0.088278  & 189.480312 	& +8.384737	    & SDSS J123758.69+082301.5 (?) & 	11.01	\\ 
iPTF14atg		    	& 0.02129		& 0.02222   & 193.186849 	& +26.470284	& IC 0831               & 10.88	\\ 
iPTF14bbr		 	    & 0.06549		& 0.06662   & 186.546284 	& +7.668036	    & SDSS J122611.21+074000.9 & 10.85	\\ 
iPTF14bdn	 	        & 0.01558       & 0.016348  & 202.687002 	& +32.761788	& UGC 8503 & 8.37	\\ 
iPTF14bpo		        & 0.07847		& 0.07838   & 258.629576 	& +31.157130	& SDSS J171429.74+310905.0 (?) & 10.77	\\ 
iPTF14bpz	 	        & 0.120$^{a}$   & 0.120     & 234.215837 	& +21.767070	& SDSS J153651.66+214556.5 & 8.48	\\ 
iPTF14bqg	 	        & 0.03291		& 0.03303   & 245.986385 	& +36.228411	& SDSS J162356.48+361339.3 & 10.84	\\ 
iPTF14ddi 	            & 0.08133		& 0.08126   & 257.639496 	& +31.659566    & SDSS J171036.45+313945.0 (?) & 11.16	\\ 
iPTF14deb		        & 0.13243		& 0.13293   & 229.614857 	& +19.742951	& SDSS J151828.02+194455.3 & 11.42	\\ 
iPTF14eje			    & 0.11888		& 0.11774   & 348.293114	& +29.191366	& SDSS J231309.15+291111.6 & 11.38	\\ 
iPTF14fpb			    & 0.061$^{a}$   & 0.060     & 11.944065 	& +11.240145	& SDSS J004746.83+111415.9 & 10.14	\\ 
iPTF14fww		        & 0.10296		& 0.10183   & 10.326226 	& +15.438180	& SDSS J004118.33+152616.2 & 10.07	\\ 
iPTF14gnl		        & 0.053727		& 0.052572  & 5.951363 	    & -3.857740 	& SDSS J002348.33-035120.6 & 10.59	\\ 
iPTF16abc (SN\,2016bln)	& 0.023196		& 0.024128  & 203.689542 	& +13.853974	& NGC 5221 & 10.85	\\ 
iPTF16auf (SN\,2016ccz)	& 0.01499		& 0.01563   & 217.788598 	& +27.236051	& MRK 0685 & 9.53	\\ 
iPTF17lf (SN\,2017lf)	& 0.01464 		& 0.01407   & 48.139952	    & +39.320608	& NGC 1233 & 10.50	\\ 
\hline
\multicolumn{6}{l}{$(a):$ Redshift determined using \texttt{SNID}. $(b):$ Redshift determined from host galaxy lines in the SN spectra. }\\
  \end{tabular}
\end{table*}

\begin{table*}
  \centering
  \caption{Fitted light curve peak magnitudes, $k$-corrected to restframe and corrected for MW extinction using the \texttt{SNooPy max\_model}.}
  \label{tab:peakmags}
  \begin{tabular}{@{}lccccccccc@{}}
    \hline \hline
SN & $B_{max}$ & $V_{max}$ & $u_{max}$ & $g_{max}$ & $r_{max}$ & $i_{max}$ & $Y_{max}$ & $J_{max}$ & $H_{max}$ \\ 
\hline
iPTF13s & -- & -- & -- & 17.47 (0.01) & 17.75 (0.01) & 18.40 (0.06) & 18.58 (0.07) & 18.22 (0.11) & -- \\ 
iPTF13ez & -- & -- & -- & -- & 17.43 (0.01) & 17.80 (0.06) & 17.62 (0.06) & 17.72 (0.08) & -- \\ 
iPTF13ft & -- & -- & -- & -- & 16.84 (0.01) & 17.50 (0.02) & 17.66 (0.05) & 17.46 (0.08) & 17.58 (0.13) \\ 
iPTF13abc & 18.29 (0.12) & 18.32 (0.09) & -- & 18.28 (0.06) & 18.27 (0.03) & 19.03 (0.05) & 19.04 (0.11) & -- & -- \\ 
iPTF13ahk & 21.02 (0.42) & 19.53 (0.39) & -- & 20.24 (0.13) & 18.66 (0.02) & 18.46 (0.10) & 17.67 (0.14) & 17.18 (0.28) & 17.91 (0.30) \\ 
iPTF13anh & 18.11 (0.03) & 17.97 (0.03) & 18.68 (0.05) & -- & 18.19 (0.01) & 18.64 (0.01) & 18.69 (0.05) & 18.58 (0.22) & -- \\ 
iPTF13aro & 19.04 (0.05) & 18.96 (0.04) & 19.35 (0.12) & 18.93 (0.06) & 19.05 (0.02) & 19.47 (0.03) & 19.38 (0.14) & 19.24 (0.31) & -- \\ 
iPTF13asv & 16.32 (0.02) & 16.37 (0.02) & 16.39 (0.04) & 16.28 (0.02) & 16.52 (0.01) & 17.23 (0.02) & 17.50 (0.04) & 17.15 (0.04) & 17.32 (0.08) \\ 
iPTF13ayw & 18.20 (0.04) & 18.18 (0.05) & -- & 18.19 (0.03) & 18.01 (0.01) & 18.50 (0.01) & 18.37 (0.06) & 18.44 (0.26) & 18.14 (0.19) \\ 
iPTF13azs & 17.93 (0.04) & 17.58 (0.04) & 18.52 (0.06) & 17.75 (0.05) & 17.60 (0.01) & 17.90 (0.01) & 17.66 (0.04) & 17.31 (0.05) & 17.13 (0.08) \\ 
iPTF13bkw & 18.39 (0.03) & 18.25 (0.03) & 18.66 (0.07) & 18.43 (0.02) & 18.29 (0.01) & 18.83 (0.06) & 18.98 (0.15) & 19.16 (0.26) & -- \\ 
iPTF13crp & 18.86 (0.02) & 18.40 (0.02) & 19.21 (0.06) & 18.70 (0.05) & 18.38 (0.01) & 18.82 (0.02) & 18.71 (0.28) & 18.26 (0.27) & -- \\ 
iPTF13daw & 19.20 (0.02) & -- & -- & 19.12 (0.02) & 19.08 (0.01) & 19.49 (0.03) & 19.14 (0.26) & -- & -- \\ 
iPTF13ddg & 18.79 (0.02) & -- & -- & 18.78 (0.01) & 18.80 (0.01) & 19.36 (0.01) & 19.22 (0.07) & 19.29 (0.26) & -- \\ 
iPTF13dge & 15.13 (0.01) & 15.19 (0.01) & 15.64 (0.03) & 15.28 (0.00) & 15.28 (0.01) & 15.86 (0.01) & 15.80 (0.04) & 15.61 (0.08) & 15.81 (0.08) \\ 
iPTF13dkj & 17.06 (0.01) & -- & -- & 16.97 (0.01) & 17.03 (0.01) & 17.53 (0.01) & 17.33 (0.05) & 17.18 (0.11) & 17.59 (0.28) \\ 
iPTF13dkx & 17.22 (0.01) & 17.04 (0.02) & 17.52 (0.04) & 17.12 (0.01) & 17.11 (0.01) & 17.53 (0.01) & 17.52 (0.07) & 17.19 (0.06) & 17.31 (0.05) \\ 
iPTF13duj & 15.11 (0.01) & 15.07 (0.01) & 15.51 (0.06) & -- & 15.09 (0.02) & 15.71 (0.02) & 15.85 (0.04) & 15.69 (0.05) & 15.89 (0.05) \\ 
iPTF13dym & 17.75 (0.02) & -- & -- & 17.50 (0.02) & 17.49 (0.02) & 17.86 (0.05) & 17.82 (0.09) & 17.78 (0.09) & 17.83 (0.14) \\ 
iPTF13dzm & 15.81 (0.02) & -- & -- & 15.59 (0.02) & 15.54 (0.01) & 15.94 (0.03) & -- & -- & -- \\ 
iPTF13ebh & 15.08 (0.01) & 14.93 (0.01) & 15.75 (0.01) & 14.95 (0.01) & 14.91 (0.01) & 15.32 (0.01) & 15.27 (0.04) & 15.01 (0.03) & 15.20 (0.05) \\ 
iPTF13efe & 18.32 (0.02) & -- & -- & 18.28 (0.02) & 18.34 (0.02) & 18.99 (0.02) & 19.13 (0.09) & 18.88 (0.33) & -- \\ 
iPTF14yw & -- & -- & -- & 16.08 (0.05) & 15.85 (0.02) & 16.43 (0.03) & 16.43 (0.12) & 16.28 (0.14) & -- \\ 
iPTF14yy & 18.43 (0.03) & 18.12 (0.05) & -- & -- & 18.12 (0.01) & 18.50 (0.01) & 18.25 (0.03) & -- & -- \\ 
iPTF14aje & 18.88 (0.06) & 18.05 (0.03) & 19.67 (0.03) & 18.71 (0.02) & 17.73 (0.01) & 17.75 (0.02) & 17.33 (0.08) & 16.76 (0.26) & 16.85 (0.38) \\ 
iPTF14ale & -- & -- & 19.33 (0.02) & 19.29 (0.01) & 19.77 (0.02) & 19.50 (0.05) & 19.59 (0.13) & 19.89 (0.14) \\ 
iPTF14bbr & -- & -- & 18.10 (0.05) & 18.25 (0.03) & 18.75 (0.08) & 18.79 (0.36) & 18.48 (0.14) & 18.79 (0.10) \\ 
iPTF14bdn & 14.78 (0.03) & 15.04 (0.03) & 14.87 (0.04) & 14.92 (0.01) & 15.45 (0.01) & 15.56 (0.04) & 15.20 (0.03) & 15.42 (0.05) \\ 
iPTF14bpo & 18.95 (0.07) & -- & -- & 18.93 (0.04) & 18.98 (0.03) & 19.43 (0.07) & 19.21 (0.10) & 19.09 (0.27) & -- \\ 
iPTF14bpz & 19.83 (0.02) & -- & -- & 19.75 (0.01) & 19.79 (0.03) & 20.43 (0.07) & 20.74 (0.29) & -- & -- \\ 
iPTF14bqg & 18.91 (0.05) & 18.30 (0.05) & -- & 18.70 (0.11) & 17.88 (0.02) & 18.07 (0.02) & 17.67 (0.07) & 16.99 (0.08) & 17.23 (0.14) \\ 
iPTF14ddi & 19.01 (0.04) & -- & -- & 18.90 (0.02) & 18.95 (0.01) & 19.50 (0.02) & -- & 19.47 (0.05) & 19.68 (0.05) \\ 
iPTF14deb & 21.00 (0.06) & 20.77 (0.04) & -- & -99.90 (-99.00) & 20.57 (0.03) & -- & -- & 20.60 (0.06) & 20.73 (0.06) \\ 
iPTF14eje & 19.42 (0.01) & -- & -- & 19.32 (0.04) & 19.54 (0.02) & 20.08 (0.08) & -- & 19.92 (0.15) & 20.10 (0.14) \\ 
iPTF14fpb & 18.14 (0.01) & -- & -- & 18.02 (0.00) & 18.11 (0.01) & 18.78 (0.02) & -- & 18.78 (0.16) & 18.93 (0.13) \\ 
iPTF14fww & -- & -- & -- & 19.40 (0.01) & 19.52 (0.06) & 20.06 (0.09) & -- & 20.08 (0.15) & 19.97 (0.11) \\ 
iPTF14gnl & 17.52 (0.04) & 17.58 (0.05) & -- & 17.53 (0.01) & 17.68 (0.02) & 18.02 (0.05) & -- & 18.07 (0.09) & 18.23 (0.13) \\ 
iPTF16abc & 15.95 (0.04) & 15.93 (0.05) & 16.07 (0.04) & 15.81 (0.01) & 15.93 (0.01) & 16.49 (0.01) & 16.67 (0.03) & 16.36 (0.03) & 16.60 (0.04) \\ 
iPTF16auf & 15.46 (0.02) & 15.47 (0.01) & 16.24 (0.12) & 15.32 (0.01) & 15.33 (0.01) & 15.70 (0.01) & 15.87 (0.05) & 15.48 (0.07) & 15.73 (0.08) \\ 
iPTF17lf & -- & -- & -- & 18.79 (0.06) & 17.27 (0.06) & 17.04 (0.03) & 16.73 (0.11) & -- & -- \\ 
\hline	
  \end{tabular}
\end{table*}

\begin{table*}
  \centering
  \caption{Light curve parameters for the supernovae using the \texttt{SNooPy color\_model}.
  }
  \label{tab:lcfits}
\begin{tabular}{@{}lccccc@{}}
\hline
SN & $T_{\rm max}$ & $s_{BV}$ & $E(B-V)_{\rm MW}$ & $E(B-V)_{\rm host}$ & $R_V$ \\
\hline \hline
iPTF13s   & 56338.13 (0.19) & 1.091 (0.027) & 0.011 & -0.011 (0.013) & 2.0 (-) \\
iPTF13ez  & 56346.87 (0.07) & 0.879 (0.020) & 0.043 & 0.309 (0.052) & 1.4 (0.2) \\
iPTF13ft  & 56356.87 (0.18) & 1.089 (0.015) & 0.015 & -0.049 (0.028) & 2.0 (-) \\
iPTF13abc & 56385.78 (1.09) & 0.850 (0.029) & 0.030 & 0.018 (0.026) & 2.0 (-) \\
iPTF13ahk & 56396.20 (0.36) & 0.508 (0.043) & 0.013 & 1.980 (0.057) & 1.1 (0.2) \\
iPTF13anh & 56414.59 (0.09) & 0.945 (0.006) & 0.022 & 0.161 (0.015) & 2.3 (0.4) \\
iPTF13aro & 56423.93 (0.20) & 0.875 (0.027) & 0.044 & 0.186 (0.019) & 2.0 (0.2) \\
iPTF13asv & 56429.54 (0.16) & 1.098 (0.018) & 0.044 & -0.030 (0.011) & 2.4 (1.2) \\
iPTF13ayw & 56431.17 (0.14) & 0.756 (0.021) & 0.029 & 0.210 (0.017) & 2.0 (0.3) \\
iPTF13azs & 56436.78 (0.15) & 1.035 (0.019) & 0.019 & 0.466 (0.013) & 3.2 (0.2) \\
iPTF13bkw & 56459.11 (0.08) & 1.014 (0.016) & 0.022 & 0.282 (0.019) & 1.0 (0.1) \\
iPTF13crp & 56527.92 (0.21) & 1.262 (0.016) & 0.050 & 0.407 (0.020) & 2.5 (0.5) \\
iPTF13daw & 56543.15 (0.56) & 0.718 (0.011) & 0.034 & 0.138 (0.021) & 3.9 (0.3) \\
iPTF13ddg & 56547.89 (0.09) & 1.014 (0.010) & 0.060 & 0.143 (0.013) & 2.5 (0.1) \\
iPTF13dge & 56556.36 (0.02) & 1.023 (0.004) & 0.078 & 0.143 (0.007) & 2.4 (0.4) \\
iPTF13dkj & 56560.45 (0.06) & 0.929 (0.011) & 0.147 & 0.167 (0.006) & 2.5 (0.5) \\
iPTF13dkx & 56565.52 (0.10) & 1.202 (0.011) & 0.027 & 0.188 (0.008) & 4.3 (0.2) \\
iPTF13duj & 56601.58 (0.11) & 1.099 (0.017) & 0.067 & 0.150 (0.012) & 1.4 (0.2) \\
iPTF13dym & 56610.27 (0.68) & 0.541 (0.042) & 0.038 & 0.021 (0.012) & 1.9 (1.5) \\
iPTF13dzm & 56614.32 (0.08) & 0.675 (0.014) & 0.049 & 0.205 (0.020) & 1.6 (0.1) \\
iPTF13ebh & 56623.24 (0.03) & 0.609 (0.005) & 0.067 & 0.069 (0.007) & 3.1 (1.0) \\
iPTF13efe & 56641.44 (0.27) & 1.193 (0.023) & 0.021 & 0.100 (0.010) & 1.7 (0.1) \\
iPTF14yw  & 56729.58 (0.16) & 0.848 (0.029) & 0.026 & 0.276 (0.019) & 1.3 (0.4) \\
iPTF14yy  & 56733.27 (0.15) & 0.802 (0.020) & 0.020 & 0.305 (0.018) & 3.2 (0.2) \\
iPTF14aje & 56758.24 (0.10) & 0.650 (0.015) & 0.152 & 0.794 (0.018) & 2.5 (0.1) \\
iPTF14ale & 56771.54 (0.26) & 0.989 (0.025) & 0.015 & 0.289 (0.019) & 1.3 (0.3) \\
iPTF14bbr & 56804.28 (0.91) & 1.028 (0.054) & 0.021 & 0.122 (0.009) & 2.4 (0.2) \\
iPTF14bdn & 56822.23 (0.09) & 1.115 (0.012) & 0.010 & 0.100 (0.006) & 3.5 (0.3) \\
iPTF14bpo & 56830.32 (0.67) & 0.757 (0.026) & 0.034 & 0.066 (0.029) & 2.0 (0.1) \\
iPTF14bpz & 56836.01 (0.22) & 1.198 (0.022) & 0.046 & 0.090 (0.018) & 2.0 (0.3) \\
iPTF14bqg & 56837.98 (0.25) & 0.910 (0.050) & 0.013 & 1.076 (0.035) & 1.4 (0.3) \\
iPTF14ddi & 56850.63 (0.10) & 0.858 (0.016) & 0.032 & 0.099 (0.023) & 1.2 (0.1) \\
iPTF14deb & 56841.45 (0.30) & 0.613 (0.034) & 0.039 & 0.222 (0.021) & 2.2 (0.4) \\
iPTF14eje & 56901.79 (0.66) & 1.084 (0.040) & 0.108 & 0.080 (0.016) & 3.0 (1.0) \\
iPTF14fpb & 56928.80 (0.53) & 1.086 (0.017) & 0.072 & 0.133 (0.017) & 1.5 (0.1) \\
iPTF14fww & 56929.64 (0.64) & 1.229 (0.035) & 0.063 & 0.081 (0.012) & 3.0 (0.2) \\
iPTF14gnl & 56957.26 (0.11) & 0.953 (0.017) & 0.027 & 0.083 (0.025) & 1.8 (0.6) \\
iPTF16abc & 57499.01 (0.08) & 1.070 (0.010) & 0.024 & 0.105 (0.005) & 2.4 (0.2) \\
iPTF16auf & 57537.80 (0.09) & 1.189 (0.012) & 0.013 & 0.248 (0.021) & 2.9 (0.3) \\
iPTF17lf  & 57776.60 (0.95) & 0.946 (0.032) & 0.134 & 2.129 (0.088) & 1.2 (0.1) \\
\hline	
  \end{tabular}
\end{table*}

\end{document}